\documentclass[12pt]{article} 
\usepackage{fullpage}
\usepackage{amsmath,amsthm,amsfonts,amssymb,amscd}
\usepackage{lastpage}
\usepackage{mathrsfs}
\usepackage{xcolor}
\usepackage{graphicx}
\usepackage{listings}
\usepackage{hyperref}
\usepackage{xfrac}
\usepackage[ruled, vlined]{algorithm2e}
\usepackage{placeins}
\usepackage[title]{appendix}
\usepackage{setspace}

\def\tX{{\tilde{X}}}
\def\tY{{\tilde{Y}}}
\def\tx{{\tilde{x}}}
\def\ty{{\tilde{y}}}
\def\Z{{\mathbb Z}}

\def\norm{{\mathcal N}}

\def\one{{\mathbf 1}}

\def\P{{\mathbf P}}

\def\D{{\mathcal D}}
\def\M{{\mathcal M}}
\def\L{{\mathcal L}}
\def\E{{\mathsf E}}

\DeclareMathOperator*{\argmin}{argmin}
\DeclareMathOperator*{\argmax}{argmax}

\usepackage[sectionbib]{natbib}
\usepackage{array,epsfig,fancyheadings,rotating}
\usepackage[]{hyperref}  
\usepackage{sectsty, secdot}
\sectionfont{\fontsize{12}{14pt plus.8pt minus .6pt}\selectfont}
\renewcommand{\theequation}{\thesection\arabic{equation}}
\subsectionfont{\fontsize{12}{14pt plus.8pt minus .6pt}\selectfont}

\usepackage{amsmath}
\usepackage{amssymb}
\usepackage{amsfonts}
\usepackage{multirow}
\usepackage{amsthm}

\newtheorem{theorem}{Theorem}

\newtheorem{proposition}{Proposition}
\theoremstyle{definition}
\newtheorem{definition}{Definition}

\begin{document}


\renewcommand{\baselinestretch}{2}



\renewcommand{\thefootnote}{}
$\ $\par


\fontsize{12}{14pt plus.8pt minus .6pt}\selectfont \vspace{0.8pc}
\centerline{\large\bf ASSESSING STATISTICAL DISCLOSURE RISK FOR}
\vspace{2pt} 
\centerline{\large\bf DIFFERENTIALLY PRIVATE, HIERARCHICAL}
\vspace{2pt} 
\centerline{\large\bf COUNT DATA, WITH APPLICATION TO THE}
\vspace{2pt} 
\centerline{\large\bf 2020 U.S. DECENNIAL CENSUS}
\vspace{.4cm} 
\centerline{Zeki Kazan and Jerome P. Reiter} 
\vspace{.4cm} 
\centerline{\it Duke University}
 \vspace{.55cm} \fontsize{9}{11.5pt plus.8pt minus.6pt}\selectfont


\begin{quotation}
\noindent {\it Abstract:}
We propose Bayesian methods to assess statistical disclosure risks for count data released under zero-concentrated differential privacy, focusing on settings with a hierarchical structure. 
We discuss applications of these risk assessment methods to differentially private data releases from the 2020 U.\ S.\ decennial census and perform empirical studies using public, individual-level data from the 1940 U.\ S.\ decennial census. In these studies, we examine how the data holder’s choice of privacy parameters affects disclosure risks and quantify increases in risk when an adversary incorporates substantial amounts of hierarchical information.

\vspace{9pt}

\par
\end{quotation}\par

\def\thefigure{\arabic{figure}}
\def\thetable{\arabic{table}}

\renewcommand{\theequation}{\thesection.\arabic{equation}}

\fontsize{12}{14pt plus.8pt minus .6pt}\selectfont

\section{Introduction}

To protect individual respondents' privacy in the 2020 decennial census public release data products, the U.S. Census Bureau 
applies a method satisfying a variant of differential privacy (\cite{abowd20222020}). 
The Census Bureau decided to use a differentially private method, which they call the TopDown algorithm (TDA), because they determined that the risks of re-identifications when using methods employed in previous censuses, namely data swapping, are too great (\cite{abowd2018}). However, some stakeholders have questioned whether or not differential privacy is necessary, for example, \cite{ruggles2019differential} and \cite{kenny2021impact}. This has led to calls from outside groups, such as \cite{jason}, for the Census Bureau to evaluate the disclosure risks of the TDA. Such risk evaluations are particularly prudent since the Census Bureau quotes an $\varepsilon = 17.91$ to generate the perturbed counts. This exceeds the recommendations of \cite{dwork2008differential}, who suggests $\varepsilon \leq \ln(3) \approx 1.10$ for typical applications, and is also likely outside the range of reasonable $\varepsilon$ suggested by \cite{lee2011much}.


In this article, we present and illustrate methods for evaluating statistical disclosure risks for differentially private count data nested in hierarchies.  
The core idea is to compute Bayesian posterior probabilities of disclosure given the released counts and assumptions about adversaries' knowledge.  We do so for counts generated with the differentially private mechanisms employed in the 2020 decennial census products with the exception that, unlike in the TDA, we do not require released counts to be non-negative or to sum consistently across geographical hierarchies; we provide rationales for this choice in Section \ref{sec:census}.  
Using data from the 1940 U.\ S.\ census, we use these methods to shed light on the potential for disclosure risks.


The remainder of this article is organized as follows. In Section \ref{sec:background}, we present relevant background on differential privacy, statistical disclosure risks, and the 2020 decennial census application. In Section \ref{sec:methods}, we describe the Bayesian methods for assessing disclosure risks, distinguishing between settings with and without hierarchical information. In Section \ref{sec:analysis}, we apply these methods to data from the 1940 decennial census using a mechanism that satisfies zero-concentrated differential privacy (\cite{ZCDP}). We illustrate how the disclosure risks vary as a function of the privacy parameters and released counts. 
Finally, in Section \ref{sec:concl}, we conclude with  future research directions.

\section{Background} \label{sec:background}

In this section, we review background material relevant to our methods, beginning with the privacy definitions that we reference throughout.

\subsection{Differential Privacy}


A mechanism $\M(\cdot)$ for releasing statistics satisfies differential privacy (henceforth, DP) if for any two data sets $D$ and $D'$ that differ in only one row,  $\M(D)$ and $\M(D')$ are ``similar.'' How similar $\M(D)$ and $\M(D')$ must be is determined by two privacy parameters, $\varepsilon \in (0, \infty)$ and $\delta \in [0, 1)$. This idea is formalized as follows, drawing from \cite{dwork2006calibrating}.


\begin{definition}[Differential Privacy]
     A mechanism $\M$ satisfies ($\varepsilon$, $\delta$)-DP if for input data sets $D,D'$ that differ in only one row and $S \subseteq Range(\M)$, $\P[\M(D) \in S] \leq e^{\varepsilon}\P[\M(D') \in S] + \delta$.
\end{definition}

\noindent When $\delta = 0$, the guarantee is referred to as pure DP, and when $\delta > 0$, the guarantee is referred to as approximate DP. 
For approximate DP, typical values of $\delta$ in practice are extremely small;  for example, the Census Bureau uses $\delta = 10^{-10}$ in their application.

With small $\epsilon$ (and $\delta$), DP provides a formal guarantee that any one person's participation in $D$ does not affect results sufficiently that an adversary could detect that person participated. In effect, this protects against an adversary who possesses information about all but one individual in $D$. 

Mechanisms that satisfy DP can have undesirably long tails. 
Because of this drawback, the Census Bureau uses a variant of DP called 
zero-concentrated differential privacy (henceforth, zCDP). The following definition draws from \cite{ZCDP}.

\begin{definition}[Zero-Concentrated Differential Privacy]
    A mechanism $\M$ satisfies $\rho$-zCDP if for inputs $D,D'$ that differ in only one row and all $\alpha \in (1,\infty)$, $D_\alpha(\M(D) ~\Vert~ \M(D')) \leq \rho\alpha$,
    where $ D_\alpha(\M(D) ~\Vert~ \M(D'))$ is the $\alpha$-R\'enyi divergence between the distributions of $\M(D)$ and $\M(D')$.
\end{definition}

\noindent zCDP allows for mechanisms where randomness is added via a Gaussian distribution, which is not possible under pure DP and has undesirable properties under approximate DP (\cite{ZCDP}).

Since DP is a stronger criterion than zCDP, any mechanism that satisfies pure DP also satisfies zCDP (see Proposition 1.4 in \cite{ZCDP}).
Converting from zCDP to DP is less straightforward; there is no guarantee that a mechanism satisfying zCDP will satisfy pure DP. Instead, for any $\delta > 0$, one can convert from zCDP to approximate DP via the following theorem (Proposition 1.3 in \cite{ZCDP}). 

\begin{theorem}\label{th:zCDP_to_DP}
    If $\M$ satisfies $\rho$-zCDP, then for any $\delta > 0$, $\M$ satisfies $(\varepsilon, \delta)$-DP for $\varepsilon = \rho + 2\sqrt{\rho\log(1/\delta)}$.
\end{theorem}

\noindent The Census Bureau uses Theorem \ref{th:zCDP_to_DP} to report the guarantees of their mechanisms in terms of approximate DP.

Finally, we define a mechanism for releasing integer-valued statistics under zCDP. Since the mechanism involves adding noise from a discrete Gaussian distribution (see \cite{canonne2020discrete}), we refer to it as the discrete Gaussian mechanism. 

\begin{definition}[Discrete Gaussian Mechanism]
    Let $x \in \Z_{\geq 0}$ be a count statistic and suppose we wish to release a noisy count $X^* \in \Z$ satisfying $\rho$-zero concentrated differential privacy. The discrete Gaussian mechanism accomplishes this by producing a count centered at $x$ with noise from a discrete Gaussian distribution with parameter $1/(2\rho)$. That is,
    \begin{equation}
        \P[X^* = x^*] = \frac{e^{-\rho(x^* - x)^2}}{\sum_{\tilde{x} = -\infty}^\infty e^{-\rho(x^* - \tilde{x})^2}}, \qquad x^* \in \Z.
    \end{equation}
\end{definition}

\noindent Notationally, $X^* \sim \textsf{DG}(x, 1/(2\rho))$. \cite{canonne2020discrete} present  an efficient algorithm  for sampling from a discrete Gaussian distribution.


\subsection{Statistical Disclosure Risk}

For a given data set, the statistical disclosure risk (henceforth used interchangeably with disclosure risk and risk) is a measure of the risk to respondent confidentiality the data holder would experience as a consequence of releasing the data set to the public (\cite{domingo2004disclosure, duncan2000bayesian}).  \cite{duncan1986disclosure} propose measuring disclosure risk by directly modeling the behavior of a Bayesian adversary with a certain target in a released data set. 
\cite{fienberg1997bayesian} present a similar approach, focusing on how the results change when bias or noise is present in the data. \cite{reiter2005estimating} extends these methods to account for the technique used to perturb the data using legacy disclosure control methods such as topcoding variables and data swapping. 

A few authors have studied  statistical disclosure risks specifically under differential privacy.  \cite{lee2011much}   use a hypothetical adversary's posterior beliefs to provide a method for selecting $\varepsilon$ for an arbitrary data set. \cite{abowd2008protective} and \cite{mcclure2012differential}  focus on the setting of generating synthetic data with binary variables. In particular, \cite{abowd2008protective} demonstrate that by measuring the disclosure risk via the posterior odds ratio between two data sets that differ in only one row, $\varepsilon$-DP provides a bound on the disclosure risk for that row.  \cite{mcclure2012differential} hypothesize a Bayesian adversary, make assumptions about what information the adversary has available a priori, and compare the adversary's posterior probability of determining the true values for a particular target to the corresponding prior probability. 

We expand on the methodology from these works, adapting the framework of \cite{mcclure2012differential} and applying it to more general settings, including categorical variables with more than two levels and data sets with a hierarchical structure, such as the 
2020 U.\ S.\ decennial census data.

\subsection{DP in the 2020 Census Data Products} \label{sec:census}

To describe the DP methods the Census Bureau is using for the 
2020 decennial census, we 
draw from \cite{abowd20222020}. We focus on the release of the PL 94-171 file, which comprises 2020 census data used for redistricting. The file includes the six summary tables listed in Table \ref{tab:cen_tabs}. Summary tables are produced across six levels of geographic hierarchy: blocks within optimized block groups, within tracts, within counties, within states, within the nation. We focus on the persons tables P1 through P5 in Table \ref{tab:cen_tabs}. 
\begin{table}[t]
\caption{\linespread{1}\normalsize PL 94-171 tables for 2020 decennial census data.}
\label{tab:cen_tabs}
\vspace{6pt}
\centering
\linespread{1}\normalsize
\begin{tabular}{l|l}
\hline
P1 & Race                                                                                        \\
P2 & Hispanic or Latino, and not Hispanic or Latino by Race                                      \\
P3 & Race for the Population 18 Years and Over                                                   \\
P4 & Hispanic or Latino, and not Hispanic or Latino by Race for \\
   & the Population 18 Years and Over \\
P5 & Group Quarters Population by Major Group Quarters Type                                      \\
H1 & Occupancy Status (Housing) \\  
\hline
\end{tabular}
\end{table}

To create these tables, the Census Bureau begins with a detailed histogram of the Group quarters, Voting age, Hispanic, and Race variables---comprising a total of 2{,}016 cells---at each location at each level of the hierarchy. We refer to this histogram as the GVHR query. The Census Bureau first generates differentially private counts (satisfying zCDP) for this histogram using the discrete Gaussian mechanism. Then, they apply a post-processing algorithm to force counts at each level of the hierarchy to sum to the counts in the level above and to ensure that no counts are negative. This post-processing starts at the national level and works down the hierarchy. 
Finally, the Census Bureau aggregates the histograms to produce the summary tables listed in Table \ref{tab:cen_tabs} and releases them to the public.

\begin{table}[t]
\caption{\linespread{1}\normalsize Allocations of $\rho$ in the 2020 decennial census. The second column displays the total proportion of $\rho$ allocated to each geographic level, and the third column displays the proportion of the $\rho$ at that level allocated to the GVHR query. Values taken from Section 8.2 of \cite{abowd20222020}.}
\label{tab:cen_rho}
\vspace{6pt}
\centering
\linespread{1}\normalsize
\begin{tabular}{l|rr|r}
\hline
Geographic Level & Total Prop. $\rho$ & GVHR Prop. $\rho$ & GVHR $\approx \rho$ \\
\hline
United States & 104/4,099 & 189/241 & 0.0509 \\
State & 1,440/4,099 & 230/4,097 &  0.0505 \\
County & 447/4,099 & 754/4,097 & 0.0514 \\
Tract & 687/4,099 & 241/2,051 & 0.0504 \\
Optimized Block Group & 1,256/4,099 & 1,288/4,099 & 0.2465 \\
Block & 165/4,099 & 3,945/4,097 & 0.0992 \\
\hline
\end{tabular}
\end{table}

The Census Bureau assigned a global $\rho$ of 2.56 that was  divided among the levels of the geographic hierarchy, as displayed in the second column of Table \ref{tab:cen_rho}. Each census block, for example, is allocated a $\rho$ of $2.56 \times 165/4{,}099 \approx 0.103$. Within each level of the hierarchy, the $\rho$ budget is further divided among 11 queries. The allocations for the GVHR query are displayed in the third column of Table \ref{tab:cen_rho}.  The remainder of the allocations can be found in Section 8.2 of \cite{abowd20222020}. The GVHR query in each census block, for example, is allocated a $\rho$ of $0.103 \times 3{,}945/4{,}097 \approx 0.099$. The approximate $\rho$ for each level is presented in the final column of Table \ref{tab:cen_rho}. The Census Bureau also reports the DP guarantee as computed via Theorem \ref{th:zCDP_to_DP}: the global $\rho = 2.56$ corresponds to $\varepsilon = 17.91$ for $\delta = 10^{-10}$. 

The Census Bureau does not release the noisy GVHR query; rather, they release 
post-processed and aggregated counts.  Several experts in DP have argued that the Census Bureau should additionally release the noisy counts without any post-processing  (e.g., \cite{boston_globe}, \cite{seeman2020private}). Doing so could allow researchers using the data to avoid biases resulting from post-processing and estimate uncertainty properly.  However, JASON, an advisory group for the U.\ S.\ government on issues related to science and technology, raises the concern that such a release could introduce disclosure risks.  In their report (\cite{jason}), they recommend that the Census Bureau should ``release all noisy measurements that are used to produce a published statistic when doing so would not incur undue disclosure risk'' (p.\ 64), and evaluate the risks that ``the released data allows an adversary to make inferences about an individual’s characteristics with more accuracy and confidence than could be done without the data released by the Census Bureau'' (p.\ 114). 
The JASON recommendations, coupled with the calls for releasing the noisy counts without post-processing, motivate our methodological developments, which we now describe.


\section{Methods for Assessing Disclosure Risks} \label{sec:methods}
\setcounter{equation}{0}

We begin with methods that 
do not utilize information from up the hierarchy, and then discuss accounting for extra information from higher levels. 

\subsection{Setting Without Hierarchical Information}\label{sec:nohierarchical}

Under DP and zCDP, the random noise protects against an adversary who possesses information about all but one individual in the data. For applications with data nested in geographic hierarchies, such as the 2020 census, this adversary seems unlikely.  The adversary would have to possess information on all but one individual in the entire United States. To assess disclosure risks, we instead consider an attack scenario where 
the adversary possesses information on all but one individual in a census block (more generally, in the lowest level of the hierarchy). For example, the adversary could be a landlord who owns all property in a block or an administrator for a group quarters institution that makes up an entire  block.  
This set of assumptions represents a type of ``worst case'' scenario; the adversary possesses the most possible information about the individuals in the block without possessing information about the target. We discuss and evaluate other attack scenarios, e.g., where the adversary knows information at the block and block-group levels, in Section S1 of the supplement.

To formalize, suppose the data comprise solely categorical variables and are organized in a hierarchy with $h$ levels. We focus on a particular group, $g_1$, at the lowest level of the hierarchy comprising  $n_1$ individuals. In the census application, $g_1$ is a census block with $n_1$ persons. The adversary possesses complete data for $n_1 - 1$ of these individuals. We seek to assess the disclosure risk for the remaining individual, henceforth referred to as individual $t$ (the targeted individual). Let $c$ be the characteristics of individual $t$, i.e., we label the combination of the variables in individual $t$'s row as $c$. Let $X_1$ be the random variable from the adversary's perspective representing the count of individuals in  $g_1$ with characteristics $c$, and let $x_{1,-t}$ be the count of individuals in  $g_1$ with characteristics $c$ excluding individual $t$; the adversary knows $x_{1,-t}$ a priori. The support of $X_1$ is $\{x_{1,-t}, x_{1,-t} + 1\}$. Let $p \in (0,1)$ be the adversary's prior probability of assigning individual $t$ to the correct category. That is,  $\P[X_1 = x_{1,-t} + 1] = p$ and 
    $\P[X_1 = x_{1,-t}] = 1-p.$
Finally, let $x_1$ be the true count of individuals in $g_1$ with characteristics $c$; here, we assume $x_1 = x_{1,-t} + 1$. The $x_1$ is unknowable to the adversary but known by the data holder (e.g., the Census Bureau).

As in the 2020 census application, we utilize zCDP and the discrete Gaussian mechanism.  Let $\eta_1$ be the added noise so that 
$\eta_1 \sim \textsf{DG}(0, 1/(2\rho_1))$. Let  $X_1^* = x_1 + \eta_1$ be the random variable representing the differentially private value of $x_1$ and $x_1^*$ be its realized outcome. The adversary's posterior probability of correctly concluding that $X_1 = x_1= x_{1,-t}+1$ is,
\begin{align}\label{eq:posterior}
    \P[X_1 = x_1 \mid X_1^* = x_1^*] &= \frac{pe^{-\rho_1(x_1^* - x_1)^2}}{pe^{-\rho_1(x_1^* - x_1)^2} + (1-p)e^{-\rho_1(x_1^* - x_{1,-t})^2}}.
\end{align}
We define the ratio of the posterior and prior probabilities as
\begin{align} \label{eq:R'}
    R'(x_1^*) = \frac{\P[X_1 = x_1 \mid X_1^* = x_1^*]}{\P[X_1 = x_1]}.
\end{align}

For the data holder, a particularly relevant measure of  disclosure risk averages \eqref{eq:posterior} over possible realizations of $X_1^*$. We have 
\begin{align} \label{eq:marginalize}
    \P[X_1 = x_1 \mid x_1 = x_{1,-t} + 1] &= \sum_{x_1^* = -\infty}^\infty \P[X_1 = x_1 \mid X_1^* = x_1^*] \nonumber \\
    & \qquad \qquad  \P[X_1^* = x_1^* \mid x_1 = x_{1,-t} + 1].
\end{align}
We condition on $x_1$ in \eqref{eq:marginalize} to emphasize the dependence of this quantity on the true value.
Taking ratios of \eqref{eq:marginalize} over the prior risk, we define
\begin{align} 
    R &= \frac{\P[X_1 = x_1 \mid x_1 = x_{1,-t} + 1]}{\P[X_1 = x_1]} 
    \nonumber \\ &= 
    \sum_{x_1^* = -\infty}^\infty \hspace{-1pt} R'(x_1^*) \, \P[X_1^* = x_1^* \mid x_1 = x_{1,-t} + 1]. \label{eq:R}
\end{align}

We also examine how the adversary can use the posterior probabilities to make decisions. Let $\hat{X}_1$ be the adversary's point estimate of $X_1$ (either $x_{1,-t}$ or $x_{1,-t}+1$). The $\hat{X}_1$ is a function of the observed noisy counts, $\D = \{ x_1^* \}$. We assume the adversary decides the value of $\hat{X}_1$ by minimizing a loss function $\L(\hat{X}_1, X_1)$. We use 
the zero-one loss, $\L(\hat{X}_1, X_1) = \one[X_1 \neq \hat{X}_1]$, where $\one[\cdot]$ is an indicator function. The Bayes estimator is then
\begin{align}
    \argmin_{\hat{X}_1} \E[\L(\hat{X}_1, X_1) \mid \D] = 
    \argmax_{\hat{X}_1} \P[X_1 = \hat{X}_1 \mid \D].
\end{align}
That is, the adversary's point estimate for $X_1$ is whichever of $x_{1,-t}$ and $x_{1,-t}+1$ has higher posterior probability. 
Since $\hat{X}_1$ is a deterministic function of $x_1^*$, the probability the adversary correctly selects $\hat{X}_1 = x_1$ is
\begin{align}
    \P[\hat{X}_1 = x_1] &= \sum_{x_1^* = -\infty}^\infty \P[X_1^* = x_1^* \mid X_1 = x_1] \nonumber
    \one\left[\P[X_1 = x_1 \mid X_1^* = x_1^*] > \frac{1}{2}\right] \\
    &= \P[X_1^* \geq \tilde{x}_1^* \mid X_1 = x_1], \label{eq:0-1_prob}
\end{align}
where $\tilde{x}_1^*$ is the smallest value of $x_1^*$ such that $\P[X_1 = x_1 \mid X_1^* = x_1^*] > \frac{1}{2}$.

\subsection{Incorporating Hierarchical Information} \label{sec:gibbs}

It makes sense that, for example, if $x_{1,-t} = 5$ and we observe that the count of people with the characteristics $c$ at one level up (the block group level) is $5$, then it should be more likely that $x_1 = 5$ than $x_1 = 6$. Thus, using hierarchical information should improve the adversary's inference. But, if the noisy count one level up is $100$, then the hierarchical information is probably not very useful in deciding between $x_1 = 5$ and $x_1 = 6$. In this section, we formalize this intuition for the case where we use two levels of the hierarchy; generalizing to more levels is straightforward conceptually.

Let $g_2$ be the group at the second level of the hierarchy containing $t$, and let $n_2$ be the number of individuals in $g_2$. In the census application, $g_2$ is the block group---comprising $n_2$ individuals---that contains census block $g_1$. Let $X_2$ be the random variable from the adversary's perspective representing the count of individuals in $g_2$ with characteristics $c$, and let $x_2$ be the true count.  Let $\eta_2 \sim \textsf{DG}(0, 1/(2\rho_2))$ be noise added via the differentially private mechanism. Let $X_2^* = x_2 + \eta_2$ be the random variable representing the differentially private version of $x_2$, and let $x_2^*$ be the realization that is released. Let $Y_1^{(1)}, \ldots, Y_1^{(d)}$ be random variables representing the counts of individuals with characteristics $c$ in the other $d$ groups at the lowest level. That is, $X_2 = X_1 + \sum_{i=1}^d Y_1^{(i)}$. Let $\eta_{1}^{(1)}, \ldots, \eta_{d}^{(1)} \overset{iid}{\sim} \textsf{DG}(0, 1/(2\rho_1))$ be the perturbations via the privacy mechanism. Set
$Y_1 = \sum_{i=1}^d Y_1^{(i)}$ and $Y_1^* = \sum_{i=1}^d (Y_1^{(i)} + \eta_1^{(i)}) = Y_1 + \sum_{i=1}^d \eta_1^{(i)}$.  The observed data is $\D = \{x_1^*, x_2^*, y_1^*\}$. 

We first briefly describe an approximation the adversary (or data holder) can make to simplify the computations. Since $Y_1^* = Y_1 + \sum_{i=1}^d \eta_1^{(i)}$, it follows that $Y_1^*$ is an approximation of $Y_1$ with noise from a sum of discrete Gaussian distributions. The noise term is difficult to handle when performing inference, so we assume that the adversary makes the approximation
\begin{align}
    \eta_1^{(i)} \overset{iid}{\sim} \textsf{DG}(0,1/(2\rho_1)) \implies \sum_{i=1}^d \eta_1^{(i)} \approx \textsf{DG}(0,d/(2\rho_1)).
\end{align}
This approximation improves as $\rho \to 0$ and $d \to \infty$ and is very accurate for the $\rho$ and $d$ used in Section \ref{sec:analysis}. See Section S2 in the supplement for details.

The adversary can use Bayesian inference to compute posterior probabilities and disclosure risks akin to those in Section \ref{sec:nohierarchical}, using the information from the second level of the hierarchy. 
Specifically, the adversary runs a Gibbs sampler to sample from the posterior distribution of $(X_1, X_2)$ given $\D$ and uses the marginal posterior for $X_1$ to estimate the posterior probability that $X_1 = x_{1,-t} + 1$.  The risk computation and point estimation techniques described in Section \ref{sec:nohierarchical} then can be applied. For the adversary, we assume the reasonable default 
choice for the prior for $X_2 \mid X_1 = k_1$ of a uniform distribution on the set $\{k_1, k_1+1, \ldots\}$.  Section S4 of the supplement discusses the sensitivity of the posterior probabilities to the choice of the prior distribution for $X_2$.

The full conditional for $X_1$ is, for $k_1 \in \{x_{1,-t}, x_{1,-t}+1\}$ and $k_1 \leq k_2$,
\begin{align}
    \P[X_1 = & ~k_1 \mid X_2 = k_2, X_1^* = x_1^*, X_2^* = x_2^*, Y_1^* = y_1^*] \nonumber \\
    &\propto \exp\left\{-\frac{d+1}{d}\rho_1\left[k_1 - \frac{dx_1^* + (k_2 - y_1^*)}{d+1} \right]^2 \right\} \, \P[X_1 = k_1]. \label{eq:X1fc}
\end{align}
The full conditional for $X_2$ is, for $k_2 \in \{k_1, k_1+1, \ldots\}$,
\begin{align}
    \P[X_2 = & ~k_2 \mid X_1 = k_1, X_1^* = x_1^*, X_2^* = x_2^*, Y_1^* = y_1^*] \nonumber \\
    &\propto \exp\left\{-\left(\rho_2 + \frac{\rho_1}{d}\right)\left[k_2 - \frac{\rho_2x_2^* + \frac{\rho_1}{d}(y_1^* + k_1)}{\rho_2 + \frac{\rho_1}{d}}\right]^2 \right\}. \label{eq:X2fc}
\end{align}
Thus, the full conditional for $X_1$ has a Bernoulli distribution, and the full conditional for $X_2$ can be easily sampled over a grid. Section S3 of the supplement includes details of the derivations.

\section{Empirical Applications with 1940 Census Data} \label{sec:analysis}
\setcounter{equation}{0}


In this section, we illustrate the methodology from Section \ref{sec:methods} on  data from the 1940 U.\ S.\ decennial census, investigating  whether the Census Bureau's choice of privacy parameters in 2020 would have led to unacceptably high disclosure risks for the GVHR query. In Section \ref{sec:1940}, we describe the 1940 decennial census data and how they differ from the 2020 decennial census data.
In Section \ref{sec:no_hier}, we consider the case where an adversary only uses the released counts from lowest level of the hierarchy. 
In Section \ref{sec:hier}, we extend to the case where the adversary leverages the released counts from a second level of the hierarchy, examining how the hierarchical information affects risks and how this effect depends on  $x_2$ and $\rho_2$. 

\subsection{The 1940 Census} \label{sec:1940}

Every 72 years, the Census Bureau is permitted to release record-level data collected in the decennial census, without any redaction for privacy protection.  Thus, the 1940 census data are an excellent testbed for our methods.

The structure of the 1940 census data differs from that of the 2020 census data.  The 1940 census data have a smaller hierarchy; the 1940 census comprises enumeration districts within counties, within states, within the country. Per the national archives, enumeration districts could be ``covered by a single enumerator or census taker in one census period that lasted several weeks'' \citep{NatArchives} and so can vary substantially in size. For illustrative purposes, we focus on a small set of enumeration districts in North Carolina that are roughly the size of the average modern-day census block.

Another important difference is the number of categories for the variables of interest. In the 2020 census, the GVHR query produces a histogram with 2,016 possible levels---the product of 2 levels for each of the Hispanic and Voting Age variables, 8 for the Group Quarters variable, and 63 for the Race variable. In 1940, this query had only 864 possible levels---the product of 2 for Voting Age, 6 for Hispanic, 8 for Group Quarters, and 9 for Race. 
The 1940 census includes the exact age of each individual, which we transform to be whether or not the person is of voting age to match  the 2020 census data releases for the PL 94-171 file.

\subsection{Setting Without Hierarchical Information} \label{sec:no_hier}

We consider enumeration district 28-21 in Dare County, North Carolina, in 1940. Table \ref{tab:28-21} displays the counts for its GVHR query.
\begin{table}[t]
\caption{\linespread{1}\normalsize Histogram for the GVHR query for enumeration district 28-21 from the 1940 decennial census.}
\label{tab:28-21}
\vspace{6pt}
\linespread{1}\normalsize
\centering
\begin{tabular}{cccc|c}
\hline
HHGQ      & VOTINGAGE         & HISPANIC     & CENRACE & Count \\
\hline
Household & Of Voting Age     & Not Hispanic & White   & 34    \\
Household & Not Of Voting Age & Not Hispanic & White   & 10    \\
Household & Of Voting Age     & Not Hispanic & Black   & 1     \\
\hline
\end{tabular}
\end{table}
This enumeration district was similar in size to a modern day census block and consisted of 45 non-Hispanic individuals residing in households. One individual has unique characteristics and so is particularly vulnerable to an attack. 

We assume the adversary possesses complete information about all the non-unique residents in enumeration district 28-21 ($g_1$) and wishes to target the resident ($t$) who is uniquely a householder of voting age, non-Hispanic, and black ($c$). 
The adversary's known count is $x_{1,-t} = 0$. We consider five adversaries who differ only in the prior probability they assign to the event that $X_1 = 1$; the first assigns probability $p = 1/2$, the second $p = 1/5$, the third $p = 1/10$, the fourth $p = 1/50$, and the fifth $p = 1/864$ (equal prior probability on the $864$ possible levels of the GVHR query in 1940).


\begin{figure}[t]
    \centering
    \includegraphics{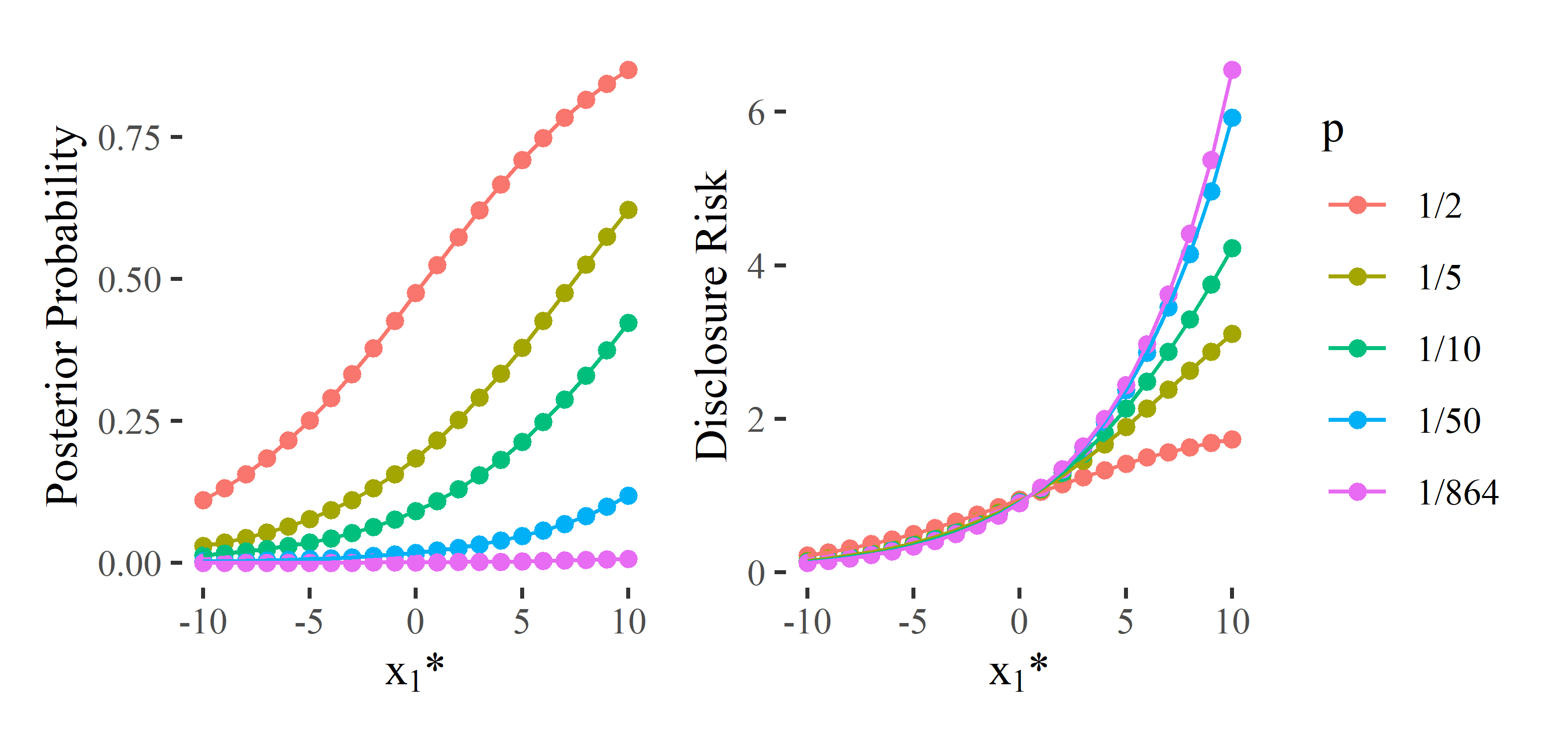}
    \caption{\linespread{1}\normalsize The left panel plots the posterior probability the adversary makes the correct decision, $\P[X_1 = 1 \mid X_1^* = x_1^*]$, as a function of $x_1^*$. The right panel plots the implied disclosure risk, $R'(x_1^*)$, as a function of $x_1^*$. The colors correspond to different adversary prior beliefs.  We set $\rho_1 = 0.099$.}
    \label{fig:plot1}
\end{figure}

We first examine the relationship between the posterior probability and the noisy released count, $x_1^*$, when $\rho_1 \approx 0.099$ as in the 2020 census application. Figure \ref{fig:plot1} displays a plot of the posterior probability the adversary assigns to the correct conclusion that $x_1 = 1$ as a function of $x_1^*$ and the implied risk $R'(x_1^*)$ for each of the five adversaries. For each adversary, both the posterior probability and the risk are monotonically increasing functions of the released  $x_1^*$. Notably, the posterior probability is greater than the prior probability $p$ (and thus, the risk greater than one) if $x_1^* \geq 1$, while the posterior is less than $p$ (and the risk less than one) when $x_1^* < 1$. This makes intuitive sense: if the released statistic is $1$ or more, this is evidence in favor of the true count being $1$, and the posterior probability that $X_1 = 1$ will increase relative to the prior probability. If the released statistic is $0$ or less, this is evidence in favor of the true count being $0$ and the posterior probability that $X_1 = 1$ will decrease relative to the prior probability.

In general terms, the risk is acceptable if it is near one, or equivalently if the posterior probability the adversary makes the correct decision is near $p$.  Thus, we should be concerned if the probability of releasing a statistic that produces a risk much greater than one is high. Table \ref{tab:tab1} presents the risks and posterior probabilities implied by several values of $x_1^*$ along with the probability of observing that value. For $x_1^* \geq 4$, the risk  substantially exceeds one, especially for adversaries with low $p$. A statistic of $x_1^* = 4$ will occur only 3.6\% of the time, and a statistic of $x_1^* = 5$ only 1.5\% of the time. While it is unlikely the released statistic will produce a risk of this magnitude in any particular census block, we will observe such risks in a sizable number of the millions of census blocks in the U.~S. Whether or not a risk of this magnitude is unacceptable is a decision for policymakers.

\begin{table}[t]
\linespread{1}\normalsize
\caption{\linespread{1}\normalsize For several values of $x_1^*$, the values of probability mass $\P[X_1^* = x_1^* \mid X_1 = 1]$, the posterior probability the adversary makes the correct decision $\P[X_1 = 1 \mid X_1^* = x_1^*]$, and the disclosure risk $R'(x_1^*)$. We set $\rho_1 = 0.099$. Posterior probabilities for $p = 1/864$ are quite small and so are omitted.}
\label{tab:tab1}
\vspace{6pt}
\centering
\begin{tabular}{cc|cccc|ccccc}
  \hline
  &  & \multicolumn{4}{c|}{Posterior Probability} & \multicolumn{5}{c}{Disclosure Risk}
  \\
$x_1^*$ & Mass & $p= \frac{1}{2}$ & $\frac{1}{5}$ & $\frac{1}{10}$ & $\frac{1}{50}$ & $\frac{1}{2}$ & $\frac{1}{5}$ & $\frac{1}{10}$ & $\frac{1}{50}$ & $\frac{1}{864}$  \vspace{1pt} \\ 
  \hline
1 & 0.161 & 0.525 & 0.216 & 0.109 & 0.022 & 1.05 & 1.08 & 1.09 & 1.10 & 1.10 \\ 
  2 & 0.119 & 0.574 & 0.252 & 0.130 & 0.027 & 1.15 & 1.26 & 1.30 & 1.34 & 1.35 \\ 
  3 & 0.073 & 0.622 & 0.291 & 0.154 & 0.032 & 1.24 & 1.46 & 1.54 & 1.62 & 1.64 \\ 
  4 & 0.036 & 0.667 & 0.334 & 0.182 & 0.039 & 1.33 & 1.67 & 1.82 & 1.96 & 2.00 \\ 
  5 & 0.015 & 0.710 & 0.379 & 0.213 & 0.047 & 1.42 & 1.90 & 2.13 & 2.37 & 2.44 \\ 
   \hline
\end{tabular}
\end{table}


\begin{figure}[t]
    \centering
    \includegraphics{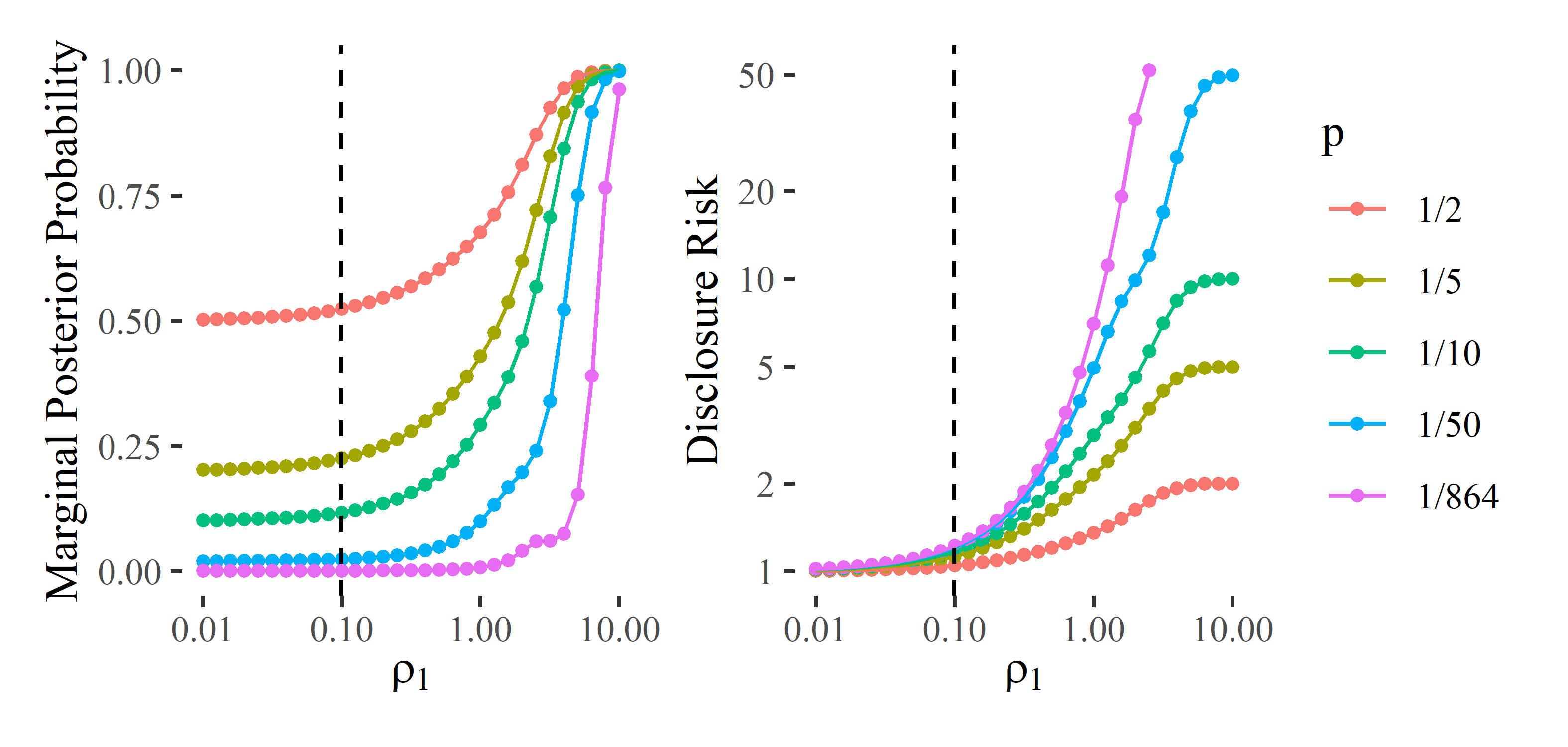}
    \caption{\linespread{1}\normalsize The left panel plots the marginal posterior probability the adversary makes the correct decision, $\P[X_1 = 1 \mid x_1 = 1]$, as a function of $\rho_1$. The right panel plots the implied disclosure risk, $R$ (defined in Equation \ref{eq:R}), as a function of $\rho_1$. The colors correspond to different adversary prior beliefs and the dashed line corresponds to $\rho_1 = 0.099$. Note the log-scales.}
    \label{fig:plot2}
\end{figure}

We next examine the expected value of the disclosure risks over realizations of the released, differentially private counts, and how 
this quantity varies with $\rho_1$. 
The first panel of Figure \ref{fig:plot2} plots the posterior probability that the adversary correctly concludes $x_1 = 1$, marginalizing over $X_1^*$, as a function of $\rho_1$. The second panel of Figure \ref{fig:plot2} plots the implied risk, $R$. As we would expect, both quantities increase monotonically with $\rho_1$, with the posterior probability increasing from $p$ at $\rho_1$ near zero to one for $\rho_1$ very large, and the risk increasing from 1 to $1/p$ along the same range. Across the various $p$, for $\rho_1 < 0.25$, the posterior probability is approximately $p$ and the risk is approximately one, indicating that the adversary gleans little from the released statistic. For $0.25 \leq \rho_1 \leq 5$, the posterior probability increases rapidly from $p$ to 1 and the risk from 1 to $1/p$, indicating that the choice of $\rho_1$ is crucial in this range.  Even small increases in $\rho_1$ can cause large increases in a hypothetical adversary's posterior probability. For $\rho_1 > 5$, the posterior probability is approximately one and the risk approximately $1/p$, indicating that parameters in this range are too high for practical settings.

\begin{table}[t]
\caption{\linespread{1}\normalsize The marginal posterior probability the adversary makes the correct decision and implied risk, $R$, for each prior probability $p$ when $\rho_1 = 0.099$.}
\label{tab:tab2}
\vspace{6pt}
\centering
\linespread{1}\normalsize
\begin{tabular}{ccc}
  \hline
$p$ & Posterior Probability & Disclosure Risk \\ 
  \hline
$1/2$ & 0.524 & 1.05 \\ 
$1/5$ & 0.225 & 1.13 \\ 
$1/10$ & 0.117 & 1.17 \\ 
$1/50$ & 0.024 & 1.21 \\ 
$1/864$ & $0.0014$ & 1.22 \\
   \hline
\end{tabular}
\end{table}

Figure \ref{fig:plot2} also demonstrates possible implications of the above discussion for the 2020 decennial census.  The Census Bureau's chosen $\rho_1 \approx 0.099$ is at the high end of the range where the posterior probability approximately equals the prior probability and the implied risk is approximately one. As shown in Table \ref{tab:tab2}, the risks are between $1.05$ and $1.21$, indicating that the Census Bureau's choice of $\rho_1$ is at a reasonable level for release of the query. If the Census Bureau were to increase $\rho_1$ by any substantial amount, the risk would increase to levels likely deemed unacceptable. On the other hand, if they were to decrease $\rho_1$ slightly, the decrease in risk would be minimal.

\begin{figure}[t]
    \centering
    \includegraphics{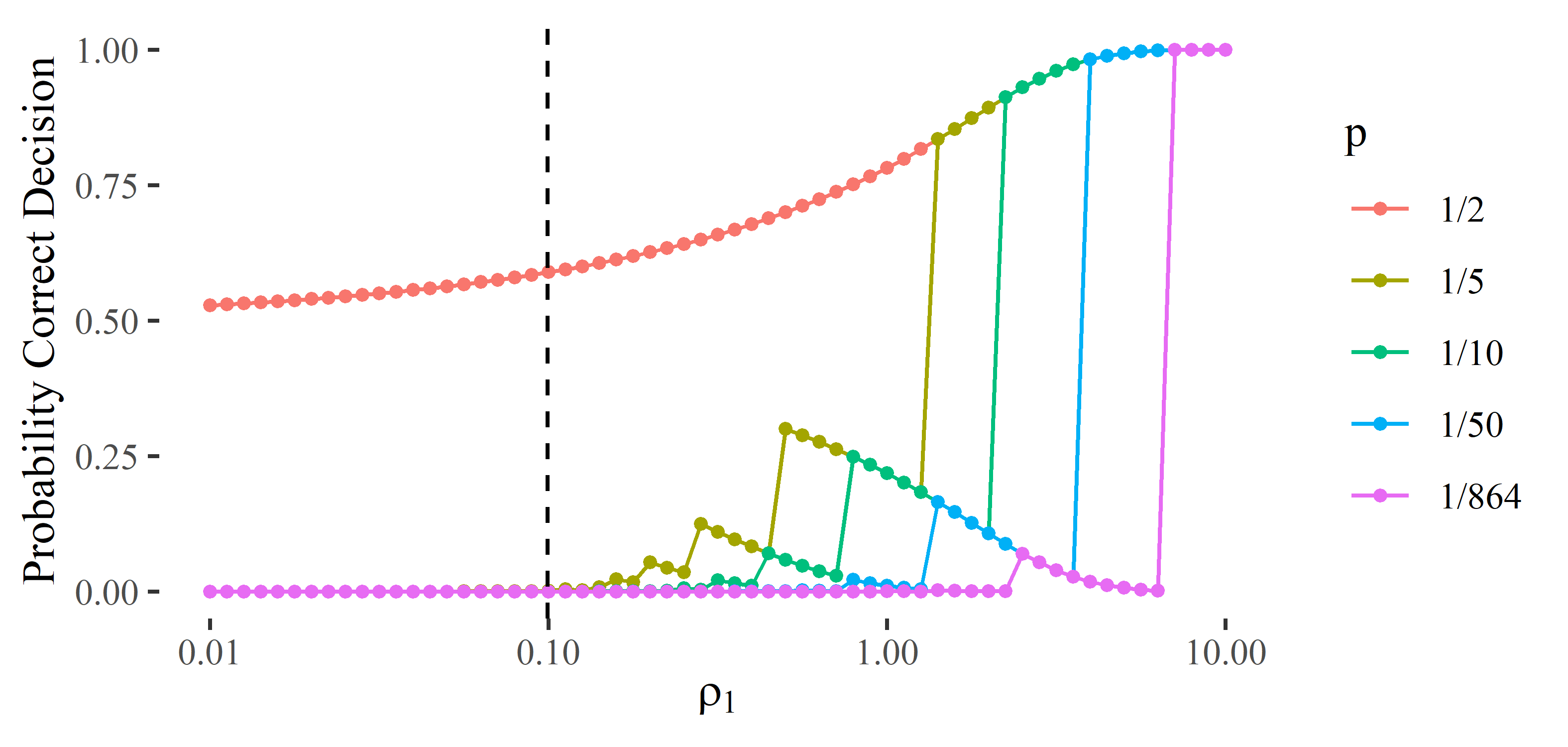}
    \caption{\linespread{1}\normalsize The probability the adversary makes the correct decision under 0-1 loss as a function of $\rho_1$. Colors correspond to different adversary prior beliefs, and the dashed line presents $\rho_1 = 0.099$. The x-axis is in log-scale.}
    \label{fig:plot3}
\end{figure}

We next examine the adversary's behavior from a decision theoretic standpoint. 
Figure \ref{fig:plot3} displays a plot of the probability that the adversary correctly selects $\hat{X}_1 = 1$ as a function of $\rho_1$ for the five adversary prior probabilities.  For $p = 1/2$, any $x_1^* \geq 1$  leads to a posterior probability greater than $0.5$.  Thus,  the curve when $p=1/2$ is simply a  plot of $\P[X_1^* \geq 1 \mid X_1 = 1]$  as a function of $\rho_1$. For $p < 1/2$ this need not be the case. For example, when $p = 1/5$ and $\rho_1 = 0.5$, $\P[X_1 = 1 \mid X_1^* = x_1^*] > 0.5$ if and only if $x_1^* \geq 2$, giving $\P[\hat{X}_1 = 1] = \P[X_1^* \geq 2 \mid X_1 = 1] \approx 0.30$. If, however, we increase to $\rho_1 = 0.6$, still $\P[X_1 = 1 \mid X_1^* = x_1^*] > 0.5$ if and only if $x_1^* \geq 2$, but now $\P[\hat{X}_1 = 1] = \P[X_1^* \geq 2 \mid X_1 = 1] \approx 0.28$. That is, since the probability of extreme values decreases as $\rho_1$ increases, the probability of making the correct decision can decrease as well. This leads to the jaggedness in the plot for $p \leq 1/5$ (and for other $p<1/2$) when $\rho_1$ is not high enough that observing $x_1^* = 1$ will give the correct decision.

As evident in Figure \ref{fig:plot3} 
the implications for the 2020 census application are very different for $p = 1/2$ and for $p \leq 1/5$. For $p = 1/2$, the probability the adversary makes the correct decision is $0.59$, which is near $0.5$. For $p \leq 1/5$, however, the probability that the adversary makes the correct decision is a fraction of a percent. Apparently, the assumptions about the adversary's prior knowledge are extremely important for this metric. An adversary who believes there is a 50-50 chance the target has characteristics $c$ will make the correct decision a substantial proportion of the time, 
but if that prior probability decreases even slightly, the  probability the adversary makes the correct decision under 0-1 loss can fall significantly.

\subsection{Incorporating Hierarchical Information} \label{sec:hier}

In this section, we explore how the incorporation of hierarchical information affects the disclosure risks. 
To do so, we consider enumeration district 39-14 in Granville County, North Carolina, one of 28 enumeration districts in the county. This district includes a white, Hispanic individual who resided in an institution for the elderly, handicapped, and poor and was not of voting age. This person was unique at both the enumeration district level (which contained 209 people) and the county level (which contained 29{,}364 people).   Thus, 
the true data is $(x_1, x_2, y_1) =  (1, 1, 0)$, where we define $y_1$ as the actual count of $Y_1$. Table \ref{tab:tab3} displays one realization of the discrete Gaussian mechanism using the Census Bureau's chosen $\rho_1 \approx 0.099$ and $\rho_2 \approx 0.246$.  The noisy counts are $\D = (x_1^*, x_2^*, y_1^*) = (2, 1, -1)$
\begin{table}[t]
\centering
\begin{minipage}[t]{.47\textwidth}
    \caption{\linespread{1}\normalsize One  sample of possible $\D = \{x_1^*, x_2^*, y_1^*\}$ with $\rho_1$ and $\rho_2$ from Census application.\label{tab:tab3}}
    \label{tab:takb3}
    \linespread{1}\normalsize
    \vspace{6pt}
    \centering
    \begin{tabular}{cc|cc|cc}
        \hline
        $x_1$ & $x_1^*$ & $x_2$ & $x_2^*$ & $y_1$ & $y^*_1$ \\
        1 & 2 & 1 & 1 & 0 & -1 \\ 
        \hline
    \end{tabular}  
    \vspace{0pt}
\end{minipage} \hspace{4mm}
\begin{minipage}[t]{.47\textwidth}
    \caption{\linespread{1}\normalsize Adversary's posterior distribution for $X_1$ given $\D$ from Table \ref{tab:tab3}, $p = 1/2$, and $10^4$ MCMC draws.}
    \label{tab:tab4}
    \linespread{1}\normalsize
    \vspace{6pt}
    \centering
    \begin{tabular}{c|cc}
        \hline
        $k_1$ & 0 & 1 \\ 
        $\P[X_1 = k_1 \mid \D]$ & 43.0\% & 57.0\% \\ 
        \hline
    \end{tabular}
    \vspace{-7pt}
\end{minipage}
\end{table}
\begin{table}[t]
\linespread{1}\normalsize
\caption{\linespread{1}\normalsize Adversary's empirical posterior distribution for $X_2$ given the data in Table \ref{tab:tab3}, $p = 1/2$, and $10^4$ MCMC draws.}
\label{tab:tab5}
\vspace{6pt}
\centering
\begin{tabular}{c|cccccccc}
  \hline $k_2$ & 0 & 1 & 2 & 3 & 4 & 5 & 6 & 7 \\ 
  $\P[X_2 = k_2 \mid \D]$ & 11\% & 40\% & 30\% & 14\% & 4.6\% & 0.7\% & 0.08\% & 0.01\% \\ 
   \hline
\end{tabular}
\end{table}

Given a $\D$, the adversary would sample from the posterior distribution of $(X_1, X_2)$. We draw 10{,}000 posterior samples from the MCMC sampler described in Section \ref{sec:gibbs}, assuming that the adversary places prior probability $p = 1/2$ on $X_1 = 1$. Table \ref{tab:tab4} summarizes the adversary's marginal posterior distribution for $X_1$.  Since $x_1^* = 2$, the posterior distribution places more weight on $X_1 = 1$ than $X_1 = 0$, but the difference is only slight because of the low $\rho$ values. If the adversary does not use hierarchical information, the posterior probability that $X_1 = 1$ is 57.4\%, indicating minimal change from using the hierarchical information in this case (in fact, the hierarchical information  slightly decreases the adversary's posterior probability of the correct choice). The adversary also can examine the posterior distribution of $X_2$, although this is of less practical interest; Table \ref{tab:tab5} presents posterior summaries. This posterior distribution places 40\% of the mass on $X_2 = 1$, and the probabilities for the right tail decline to zero relatively quickly. 

For the remainder of this section, we consider the disclosure risks at different values of $\D$. 
We assume throughout that the adversary sets $p = 1/2$. We focus on the case where the targeted individual is unique at the lowest level of the hierarchy; see Section S1.1 of the supplement for discussion of extending to non-unique individuals.


\begin{figure}[t]
    \centering
    \includegraphics{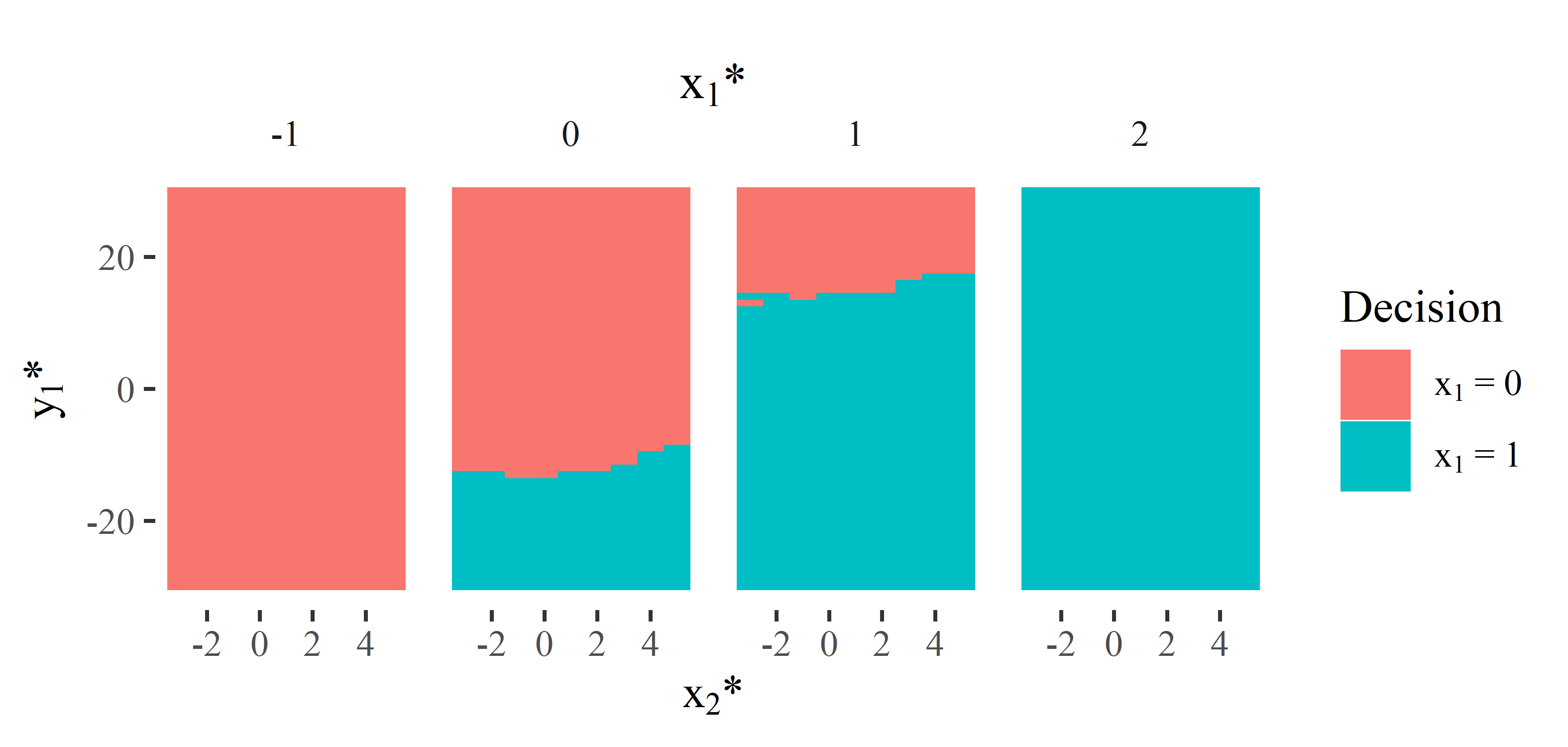}
    \caption{\linespread{1}\normalsize Adversary's decision under 0-1 loss for each combination of $x_1^*$, $x_2^*$, and $y_1^*$. Privacy parameters are set as in the census application, $p = 1/2$, and $d = 27$. $10^3$ MCMC draws are taken for each combination in most cases. When the posterior probability $x_1 = 1$ is close to 0.5, the number of MCMC draws is increased to $2.5 \times 10^5$.}
    \label{fig:plot4}
\end{figure}

We begin by examining when the hierarchical information affects the adversary's decision for $\rho_1 \approx 0.099$ and $\rho_2 \approx 0.247$, as in the 2020 decennial census. To do so, we enumerate all reasonable combinations of $x_2^*$ and $y_1^*$, which we choose as   
$-3 \leq x_2^* \leq 5$ and $-30 \leq y_1^* \leq 30$; this region contains over 99\% of the probability mass. Figure \ref{fig:plot4} displays the decisions for these combinations.  Without hierarchical information, the adversary always decides that $x_1 = 1$ when $x_1^* \geq 1$ and $x_1 = 0$ when $x_1^* \leq 0$.  Using the hierarchical information does not change the adversary's decision when $x_1^* \geq 2$ or $x_1^* \leq -1$. But when $x_1^* \in \{0,1\}$, the hierarchical information can change the adversary's decision. When $x_1^* = 1$ and $y_1^*$ is positive and large, the adversary decides that $x_1 = 0$, whereas they choose $x_1 = 1$ without the hierarchical information. Similarly, when $x_1^* = 0$ and $y_1^*$ is negative and large in absolute value, the adversary  decides that $x_1 = 1$, whereas they choose $x_1 = 0$ without the hierarchical information. Exactly how large $y_1^*$ must be depends on the observed $x_2^*$. The tiles on Figure \ref{fig:plot4} where the hierarchical information causes the adversary to correctly change their decision correspond to 2.48\% of the probability mass, while the tiles where the reverse occurs correspond to 1.83\% of the probability mass. Thus, the adversary has a slight net gain from the hierarchical information, increasing the probability of a correct decision by 0.65\%. 

\begin{figure}[t]
    \centering
    \includegraphics{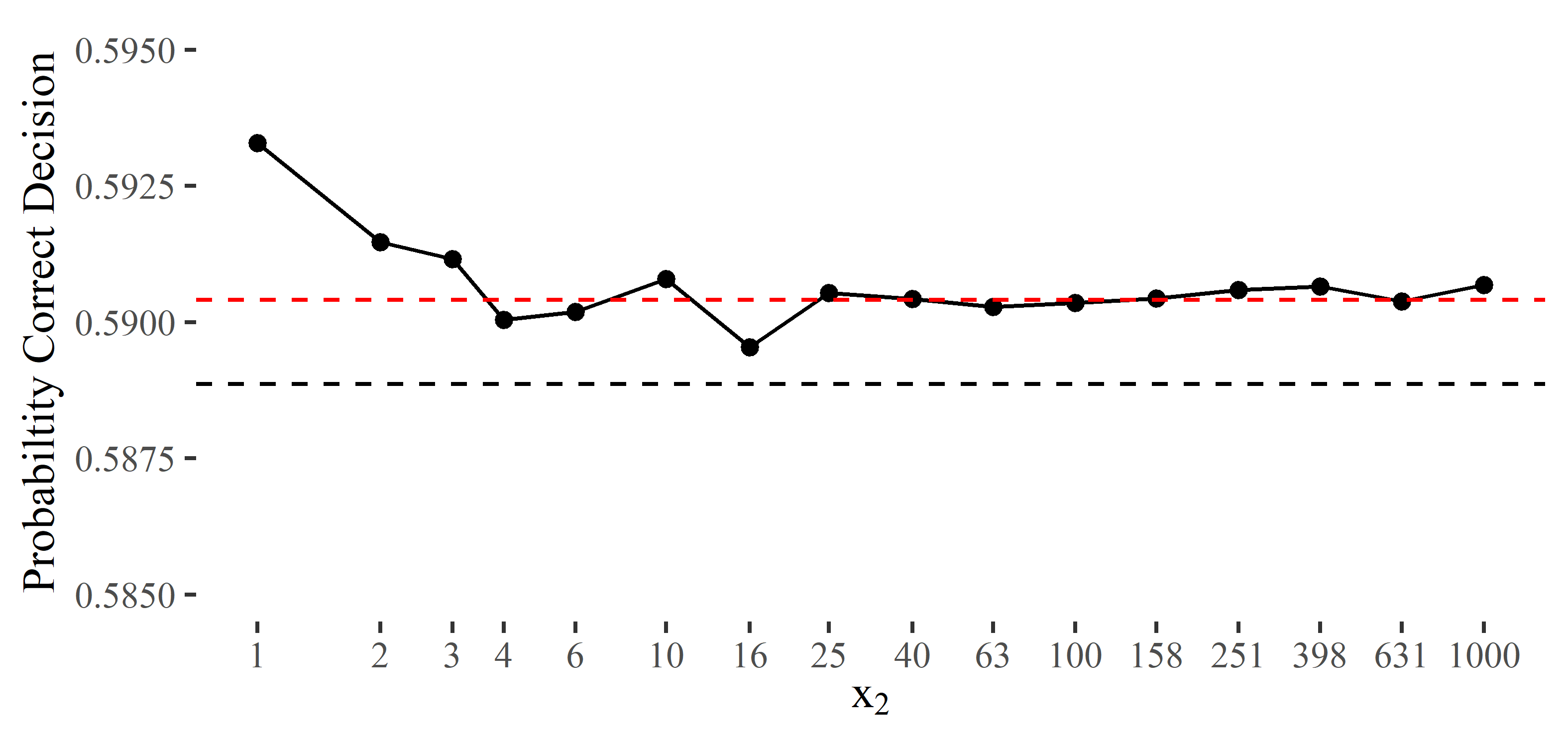}
    \caption{\linespread{1}\normalsize Proportion of times adversary correctly concludes that $x_1 = 1$ as a function of $x_2$. Proportions are over $10^6$ random draws of $\D$. For each draw, $10^3$ MCMC samples are used to estimate the posterior. The dashed black line is the corresponding probability when ignoring hierarchical information, and the dashed red line is the average proportion over $x_2 \geq 4$. We set $\rho_1 = 0.099$, $\rho_2 = 0.247$, $p = 1/2$, and $d = 27$.}
    \label{fig:plot5}
\end{figure}

We next examine how sensitive the disclosure risks are to the true count at the second level, $x_2$. 
How would the results change if the target was unique  at the lowest level, but not at the second level? To examine this, we 
alter the true counts, adding individuals to group $g_2$ with characteristics $c$. Figure \ref{fig:plot5} displays the proportion of times the adversary makes the correct decision as a function of $x_2$.
Of note, the y-axis has a small range; even with $10^6$ samples per point the Monte Carlo noise obscures the relationship. Regardless, $x_2 = 1$ is a clear outlier; the probability the adversary makes the correct decision is not only higher than the corresponding probability when ignoring hierarchical information, but higher than the analogous probabilities for $x_2 > 1$. 
For $x_2 > 3$, the probabilities of making the correct decision are centered around 59.05\% with a small amount of Monte Carlo error.

Applying these findings to the 2020 census application, the main implication is that the conclusions when $x_2 = 1$ generalize to $x_2 > 1$, although the increase in the probability the adversary makes the correct decision is less pronounced. This means that the discussion about the trade off between privacy and accuracy in terms of the Census Bureau's choice of $\rho_2$ applies no matter the value of $x_2$. A remarkable feature of Figure \ref{fig:plot5} is how small a difference is made by changes in $x_2$.


\begin{figure}[t]
    \centering
    \includegraphics{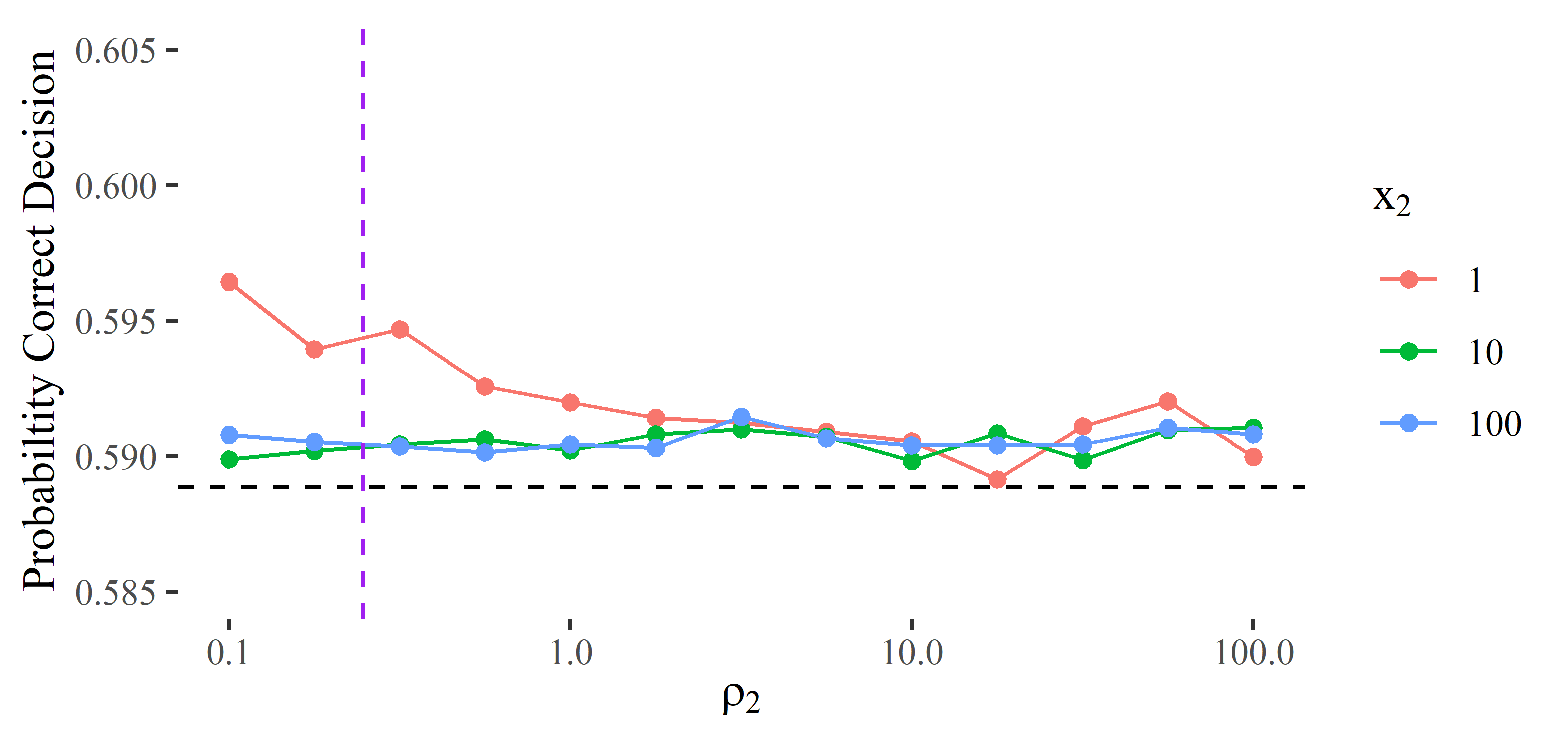}
    \caption{\linespread{1}\normalsize Plot of the proportion of the time the adversary correctly concludes that $x_1 = 1$ as a function of $\rho_2$, colored for a few selected $x_2$. The proportions are over $5 \times 10^5$ MC draws for $\D$. For each draw, $10^3$ MCMC samples are used to estimate the posterior. The dashed black line is the corresponding probability when ignoring hierarchical information and the dashed purple line is $\rho_2 = 0.247$. We set $\rho_1 = 0.099$, $p = 1/2$, and $d = 27$.}
    \label{fig:plot6}
\end{figure}

Finally, we examine the effect of changes in the second level privacy parameter, $\rho_2$. Figure \ref{fig:plot6} displays the probability the adversary makes the correct decision as a function of $\rho_2$ for $\rho_1 \approx 0.099$ and a few values of $x_2$. The effect of $x_2 = 1$ observed previously---increasing the adversary's probability of making the correct decision from when $x_2 > 1$---lessens as $\rho_2$ increases. For $\rho_2 > 1$, the probabilities for $x_2 \in \{1, 10, 100\}$ are quite similar. The most striking feature of Figure \ref{fig:plot6} is that even for extremely large values of $\rho_2$, the probability does not noticeably increase for any of the selected $x_2$. When ignoring hierarchical information, the corresponding probability is 58.89\%.  Even when $\rho_2 = 100$ the probability only increases to 59.2\%.  This finding is a feature of the specific attack scenario.  Because the adversary knows only $x_{1,-t}$ and not the counts in other blocks, 
knowing $x_2$ nearly exactly does not tell them 
which blocks making up $g_2$ contain the individuals with characteristics $c$. Thus, their probability of making the correct decision with respect to block $g_1$ improves as the uncertainty in $X_2$ decreases, but does not go to $1$. Mathematically, in (\ref{eq:X2fc}) as $\rho_2 \to \infty$, $\P[X_2 = x_2 \mid X_1 = k_1, \D] \to 1$, causing the full conditional for $X_1$ in (\ref{eq:X1fc}) to converge to a constant function of $x_2$ (but not to $1$ unless $\rho_1 \to \infty$). In the supplement, we show that hierarchical information can increase risks more noticeably in other attack scenarios, e.g., when the intruder knows  $g_2$.

Overall, the results suggest that the adversary gains little from using the hierarchical information in this attack scenario.  This may provide  evidence that the Census Bureau's choice of $\rho_2$ is reasonable from a disclosure risk perspective.  It also suggests that 
the Census Bureau could increase $\rho_2$---and likely the $\rho$ values at  higher levels of the hierarchy---without significantly compromising the disclosure risk, at least under this attack scenario. This would mean that more accurate statistics could be released at the upper levels of the hierarchy. Essentially, if we assume the adversary only has complete information about everyone except the target at the lowest level of the hierarchy, then only the choice of $\rho_1$ has a meaningful effect on the probability the adversary makes the correct decision. 

\section{Conclusion} \label{sec:concl}
\setcounter{equation}{0}

We provide methodology to compute statistical disclosure risks for categorical data with many levels released under zCDP, while incorporating hierarchical information. 
Following the suggestion in the JASON report, we demonstrate how to conduct empirical analyses that could be used to evaluate the effect on  disclosure risks of releasing the GVHR query at the census block level prior to post-processing. In our studies with 1940 census data, we find that, when assuming the adversary possesses information about all but one individual at the lowest level of the hierarchy, the main factor affecting the disclosure risk is $\rho_1$, the privacy parameter at the lowest level. The hierarchical information does not have an appreciable affect on the accuracy of the adversary's posterior inference under these assumptions. 

The redistricting files are only one set of counts from the 2020 census released by the Census Bureau; others are released over time.  Thus, from the lens of differential privacy, it is reasonable to ask about the value of assessing disclosure risks for specific attack scenarios at a point in time.  We believe such assessments have a useful role to play.  First, at the stage of algorithm design, they can help the data holder, including decision makers who may comprehend Bayesian probabilities more readily than bounds on R\'enyi divergences, understand the risks inherent in different choices of privacy parameters. Related, they can help the data holder explain the privacy protection to the public, as 
posterior probabilities and posterior-to-prior ratios can be more interpretable than guarantees expressed in terms of privacy parameters (\cite{hotz2022balancing}). 
We also note 
that the risk measures 
have desirable composition properties in settings where the same attack is applied to sequential releases, with the total disclosure risk from $m$ releases being equivalent to the product of the risk from each release (see Section S5 of the supplement for details). 
We note, however, that settings where the output of one release is used as side information for a different attack strategy are less straightforward; analysis of the risk composition in such settings is an area for future research.

Naturally, our findings are specific to a particular attack scenario---an adversary with complete information about everyone in $g_1$ except the target, about whom they know nothing. We can modify these assumptions in a number of ways 
and still use the same approach; see Section S1.2 of the supplement for a few examples. If we make large changes to the adversary's assumed knowledge, however, the results of the analysis may change. Section S1.3 of the supplement includes an example of this, whereby we assume the adversary knows the target is unique at multiple levels of the hierarchy. Here, the effect of the hierarchical information can be stronger. An adversary with this type of knowledge may or may not be realistic in some settings; whether this is the case for the decennial census is a decision for the Census Bureau.

As noted previously, we do not consider the TopDown algorithm's post-processing step or population invariants. The addition of a post-processing step by itself, i.e., absent invariants derived directly from the confidential data, does not affect the formal privacy guarantee and also should not increase the statistical disclosure risks. 
Either the post-processing step is invertible, in which case the risk analysis does not change, or it is not invertible, in which case the computation of disclosure risks is far more uncertain and computationally difficult \citep{gong:meng:20}. We illustrate this in Section S6 of the supplement. The population invariants include the total populations of each state, total number of housing units in each census block, and number of occupied group quarters of each type in each census block (\cite{abowd20222020}). It is unclear how possession of these quantities would affect our disclosure risk measures, since they are not easily related to the counts an adversary considers in our methods. Future work could provide a more formal analysis of these points, examining in more detail how much extra protection could be offered by the post-processing step (it is highly unlikely to be invertible) and how adversaries could utilize population invariants in their prior or data distributions.

This work also points to directions for future research. One direction involves relaxing the assumption that the adversary possesses complete information about all but one individual. For example, it may be possible to adapt methods used in \cite{mcclure2016assessing} to examine inferences when the adversary possesses information about all but two---or all but $n$---individuals. Another direction involves using disclosure risk measures to approximate an ``empirical'' DP bound for a data set released under DP or zCDP. For the 2020 decennial census, the Census Bureau quotes a total $\varepsilon = 17.91$, 
computed via composing the $\rho$ allocations at the six levels of the hierarchy and using Theorem \ref{th:zCDP_to_DP}.
Our results indicate that reporting the privacy guarantee in this way may understate the degree of privacy. Other works, for example the partial DP of \cite{ghazi2022living}, examine how to produce a more meaningful parameter for interpretation when the $\varepsilon$ from DP is large. A method based on disclosure risks  also may be possible and useful in practice.

\section*{Supplementary Materials}

The online supplement contains a proposition on the sums of discrete Gaussians, derivation of the full conditionals, and analysis of prior sensitivity. It also contains discussion of extensions to the methods and analysis that incorporate other attack scenarios, post-processing, and sequential releases.
\par
\section*{Acknowledgements}

This research was supported by NSF grant SES-2217456.
\par


\bibhang=1.7pc
\bibsep=2pt
\fontsize{9}{14pt plus.8pt minus .6pt}\selectfont
\renewcommand\bibname{\large \bf References}
\expandafter\ifx\csname
natexlab\endcsname\relax\def\natexlab#1{#1}\fi
\expandafter\ifx\csname url\endcsname\relax
  \def\url#1{\texttt{#1}}\fi
\expandafter\ifx\csname urlprefix\endcsname\relax\def\urlprefix{URL}\fi

  \bibliographystyle{chicago}      
  \bibliography{bib}   







\vskip .65cm
\noindent
Zeki Kazan, Duke University
\vskip 2pt
\noindent
E-mail: zekican.kazan@duke.edu
\vskip 2pt

\noindent
Jerome Reiter, Duke University
\vskip 2pt
\noindent
E-mail: jreiter@duke.edu

\normalsize
\newpage
\vspace{.55cm}
 \centerline{\bf Supplementary Material}
\vspace{.55cm}
\fontsize{9}{11.5pt plus.8pt minus .6pt}\selectfont
\noindent
This supplementary material contains a proposition on the sums of discrete Gaussian random variables, derivations of the full conditionals for the Gibbs sampler in the main text, and discussion of extensions of targeting non-unique individuals and other types of disclosure attacks.

\par

\setcounter{section}{0}
\setcounter{equation}{0}
\def\theequation{S\arabic{section}.\arabic{equation}}
\def\thesection{S\arabic{section}}

\fontsize{12}{14pt plus.8pt minus .6pt}\selectfont

\section{Other Attacks}
\setcounter{equation}{0}

This section examines other attacks an adversary could perform in addition to the attack focused on in the main text. One class of attacks considers the same attack as in the main text, but with a target who is not unique in group $g_1$. Another class considers different attacks that are mathematically equivalent to the attack from the main text. A final class considers an attack by an adversary with substantially more information than the adversaries examined in the main text.


\subsection{Non-unique Individuals} \label{sec:oth_att1}

The empirical analysis in Section 4 of the main text focuses on the case where the targeted individual is unique at the lowest level of the hierarchy (the block-level in the 2020 decennial census application).  This assumption is not necessary for the methodology. If rather than $x_{1,-t} = 0$ and $x_1 = 1$, we had $x_{1,-t} = m$ and $x_1 = m+1$---i.e., the adversary knows there are $m$ individuals other than the target with characteristics $c$---the analysis would be identical to what is presented in the main text. All the probabilities plotted in Section 4 of the main text would stay the same, and all plots involving $X_1^*$ and $X_2^*$ would be shifted by $m$. 

\subsection{Equivalent Attacks} \label{sec:oth_att2}

The methodology and empirical evaluations in the main text consider the scenario where an adversary is interested in determining whether individual $t$ has characteristics $c$.  In this section we describe how several other attacks map onto our notation with only the meaning of the prior probability, $p$, changing. The results from the main text thus can be applied to these other attacks directly.

An adversary may seek to determine whether the target filled out the census at all. In this setting, we assume the adversary possesses the complete information for all $n_1$ individuals in $g_1$, and believes a priori with probability $p_f \in (0,1)$ that individual $t$ actually filled out the census. The adversary assumes that the other $n_1-1$ individuals filled out the census accurately. Thus, the data holder can simply replace $p$ with $p_f$ in the main text and examine the risk from this attack.


Another adversary may seek to determine whether a census respondent lied or made a mistake when completing the census. In this setting, we assume the adversary possesses complete information for all $n_1$ individuals in $g_1$, and believes a priori with probability $p_{\ell} \in (0,1)$ that individual $t$ reported the correct information. The adversary assumes that the other $n_1 - 1$ individuals filled out the census accurately. Thus, the data holder can simply replace $p$ with $p_\ell$ in the main text and examine the risk from this attack.

Finally, an adversary may seek to determine an unknown variable. In this setting, we assume the adversary possesses the complete information for all $n_1$ individuals in $g_1$, except that they do not know one of the variables for individual $t$. For example, the adversary could know individual $t$'s race, HHGQ status, and whether they are of voting age, but not their ethnicity. Let $c_e$ be the true ethnicity of the individual as reported on the decennial census, and let $p_v$ be the  prior probability the adversary assigns to individual $t$ having ethnicity $c_e$. The data holder can simply replace $p$ with $p_v$ in the main text and examine the risk from this attack.


\subsection{Adversaries with Additional Information}

The settings in Sections \ref{sec:oth_att1} and \ref{sec:oth_att2} presume the adversary only has information on individuals in $g_1$.  Another class of attacks presumes the adversary has information at higher levels of the hierarchy as well.  For example, and as suggested by a reviewer, consider an adversary who seeks to determine an unknown variable for target $t$, say the individual's ethnicity.  Let $c_e$ be the true ethnicity of the target (unknown to the adversary), and let $c_{-e}$ be the true characteristics of the target for the other three variables (known to the adversary). We now include the additional assumption that the adversary knows a priori that the target's value of $c_{-e}$ is unique at $\ell$ levels of the hierarchy. For example, if $\ell = 2$, the target is the only individual in their block group with characteristics $c_{-e}$; if $\ell = 4$, the target is the only individual in their county with characteristics $c_{-e}$. Because the target is so distinct, the released noisy counts $X_{1}^*, \ldots, X_{\ell}^*$ can be combined to improve the adversary's posterior, which now has the form
\begin{align}
    \P[X_1 = x_1 &\mid X_1^* = x_1^*, \ldots, X_\ell^* = x_\ell^*] \nonumber \\
    &= \frac{p \prod_{i=1}^\ell e^{-\rho_i(x_i^* - x_1)^2}}{p \prod_{i=1}^\ell e^{-\rho_i(x_i^* - x_1)^2} + (1-p) \prod_{i=1}^\ell e^{-\rho_i(x_i^* - x_{1,-t})^2}}.
\end{align}
We can marginalize over the $\ell$ noisy counts, as in the main text, to compute the marginal posterior the adversary makes the correct conclusion and the corresponding disclosure risk. 

\begin{figure}[t]
    \centering
    \includegraphics{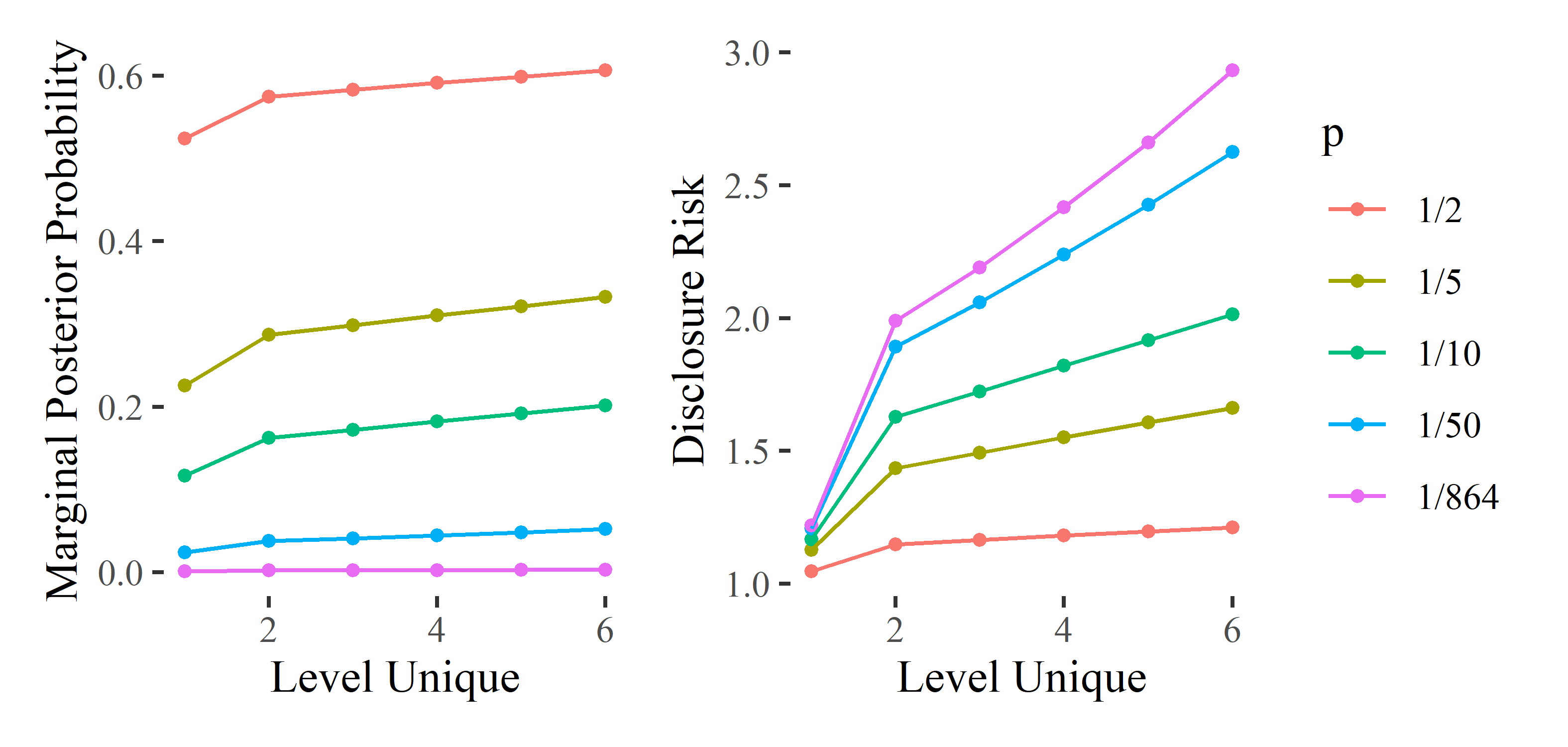}
    \caption{The left panel plots the marginal posterior probability the adversary makes the correct decision that $X_1=1$ as a function of the number of levels of the hierarchy at which the target is known to be unique. The right panel plots the corresponding implied disclosure risk. The colors correspond to different adversary prior beliefs; the $\rho$ at each level is the value used by the U.~S. Census Bureau in 2020.}
    \label{fig:plot11}
\end{figure}

Figure \ref{fig:plot11} plots the marginal posterior and disclosure risk as a function of $\ell$ for adversaries with different prior beliefs, where the prior parameters are set as in the analysis in the main text (levels 3-6 all have $\rho_i \approx 0.05$; see Table 2 in the main text for details). As expected, the risk increases as a function of $\ell$ due to the increasing amount of information available to the adversary. For all priors, the increase is most substantial between $\ell = 1$ and $\ell = 2$, since $\rho_2$ is the largest privacy parameter and thus provides the most accurate release. Overall and in contrast to the findings from the main text, we conclude that, for this type of attack, the hierarchical information can sharpen the adversary's estimates substantially. 

To carry out this attack, this adversary requires detailed information across geographical hierarchies. Whether this is a realistic adversary or not is a matter for policymakers to determine in their particular scenarios.

\section{Sums of Discrete Gaussian Random Variables}\label{app:DG_approx}
\setcounter{equation}{0}

This section focuses on the following proposition.
\begin{proposition}
    Let $Z_1, \ldots, Z_n \overset{iid}{\sim} \textsf{DG}(0, s = 1/(2\rho))$. Then, for $\rho < 1$ and $n$ large, $\sum_{i=1}^n Z_i$ is well approximated by $\textsf{DG}(0, s = n/(2\rho))$.
\end{proposition}

\noindent We present an informal proof of this fact, based on empirical results.

\begin{figure}[t]
        \centering
        \includegraphics[width=\columnwidth]{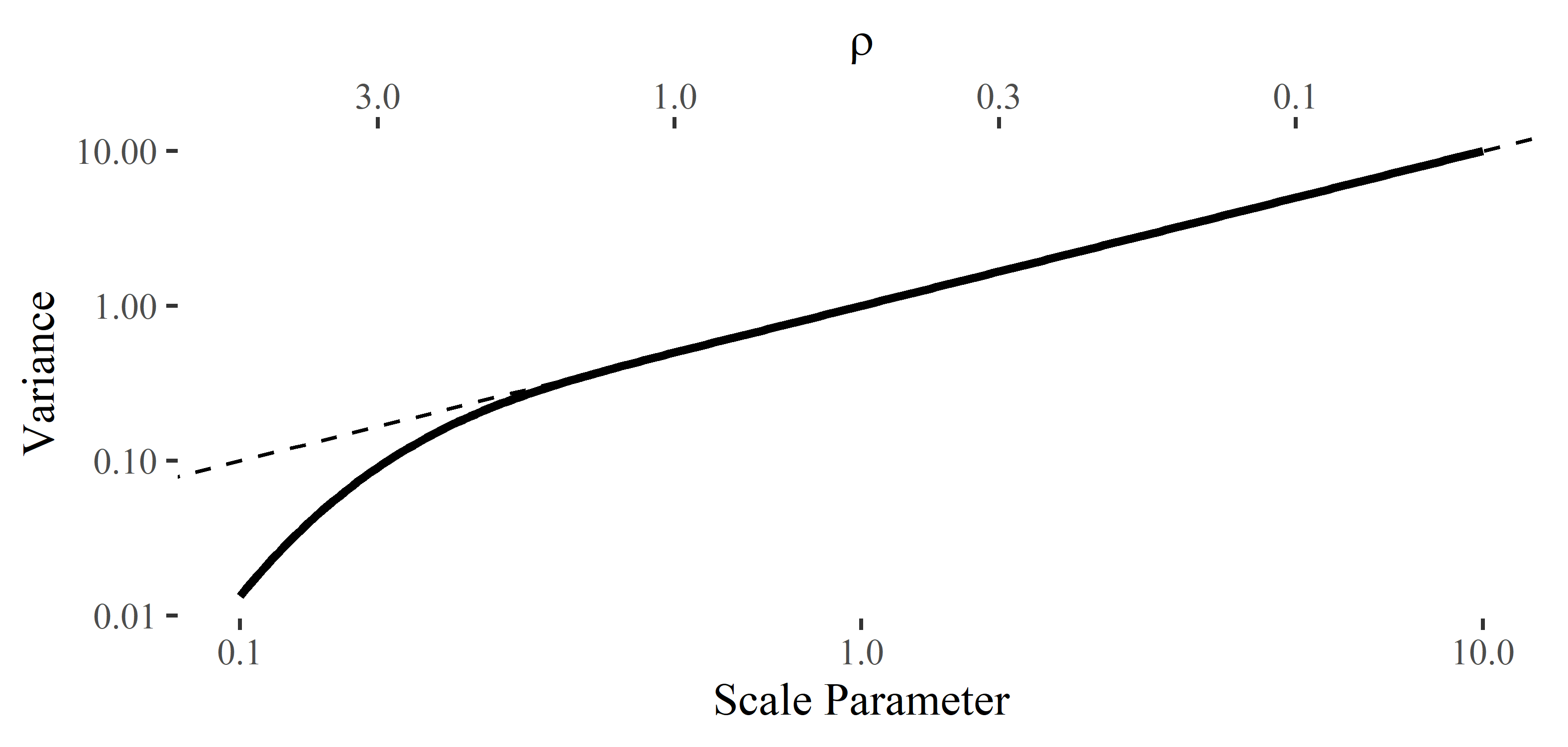}
        \caption{\linespread{1}\normalsize Plot of $\sigma^2$, the variance of $Z_i$ (computed to very high precision), as a function of the scale parameter of the discrete Gaussian, $s$. The dashed line is the line $\sigma^2 = s$; the upper axis presents $\rho = \frac{1}{2s}$ for comparison. Note the log scales.}
        \label{fig:plot7}
    \end{figure}

    To begin, we denote the variance of each $Z_i$ as $\sigma^2$. Figure \ref{fig:plot7} plots $\sigma^2$ as a function of both the scale parameter, $s$, and $\rho = 1/(2s)$. We see that for $\rho < 1$, which corresponds to $s > 0.5$, the approximation $\sigma^2 \approx s = 1/(2\rho)$ is quite accurate. Empirically, $|\sigma^2 - s| < 0.002$ for all $s > 0.5$ and $|\sigma^2 - s| < 10^{-6}$ for all $s > 1$. Thus, we are able to approximate the variance of the $Z_i$ in this range of $\rho$ with $s$.
    
\begin{figure}[h!]
    \centering
    \includegraphics[width=\columnwidth]{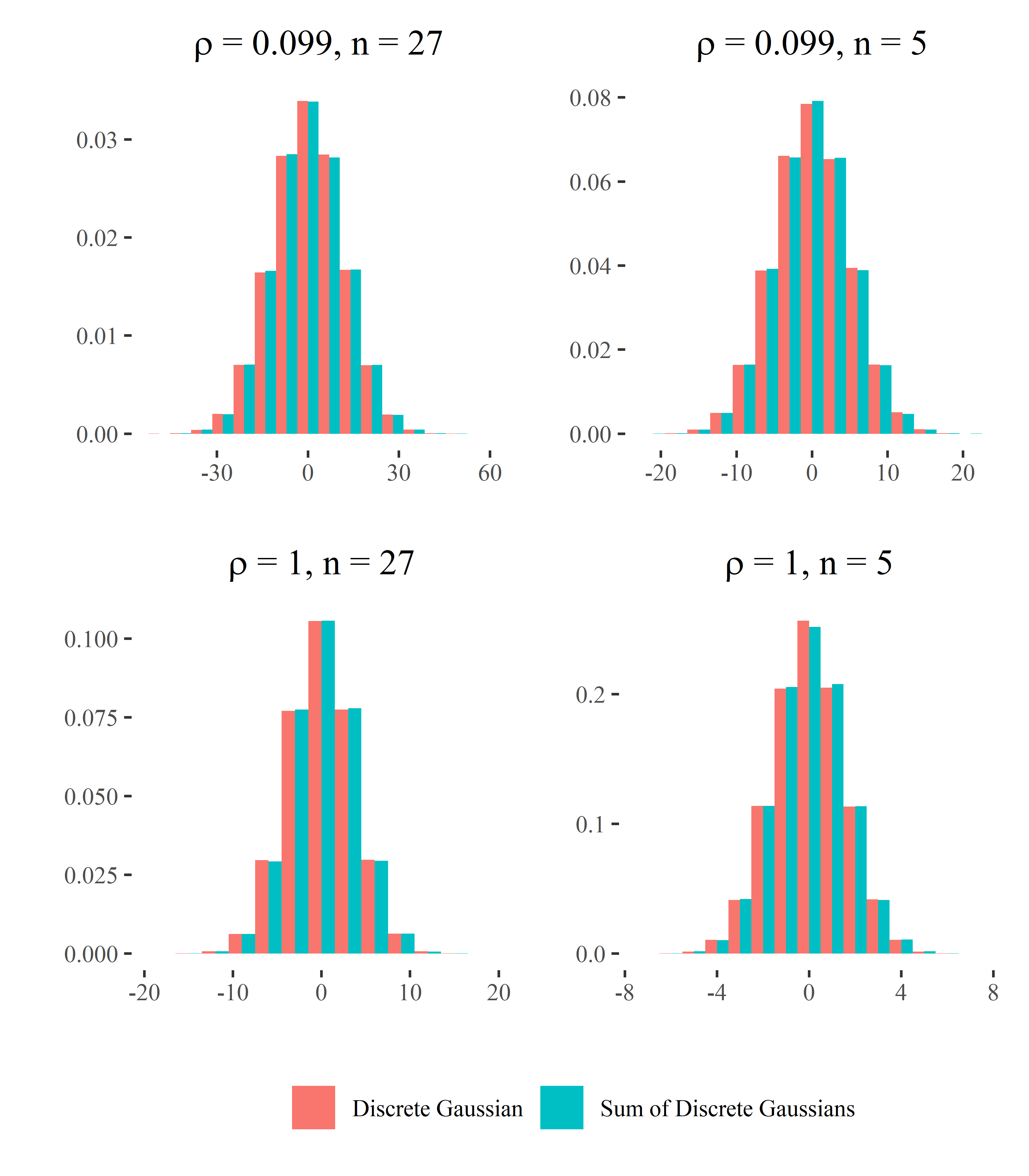}
    \caption{\linespread{1}\normalsize Histograms comparing the probability mass function of $\textsf{DG}(0, n/(2\rho))$ (in red) to the probability mass function of $\sum_{i=1}^n Z_i$ (in blue). Histograms are included for $\rho = 0.099$ on the top row, $\rho = 1$ on the bottom row, $n = 27$ on the left column, and $n = 5$ on the right column.}
    \label{fig:plot8}
\end{figure}

    For $n$ sufficiently large, we can apply the Central Limit Theorem, which gives the approximation
    \begin{equation}
        \sum_{i=1}^n Z_i \approx \norm(0, n\sigma^2).
    \end{equation}
    As this distribution is discrete and $n\sigma^2 = n/(2\rho) \gg 1$, it makes sense intuitively to instead use the approximation,
    \begin{equation}
        \norm(0, n\sigma^2) \approx \textsf{DG}(0, n\sigma^2) = \textsf{DG}\left(0, \frac{n}{2\rho}\right).
    \end{equation}
    Combining the two approximations gives
    \begin{equation}
        \sum_{i=1}^n Z_i \approx \textsf{DG}\left(0, \frac{n}{2\rho}\right).
    \end{equation}
    This approximation is quite accurate in practice. Figure \ref{fig:plot8} compares the probability mass function of $\textsf{DG}(0, n/(2\rho))$ to $\sum_{i=1}^n Z_i$ for $\rho \in \{1, 0.099\}$ and for $n \in \{5, 27\}$. We note that 0.099 is the value of $\rho_1$ used  in the 2020 census application, and $n = 27$ is used in Section 4.3 of the main text. The approximation does extremely well for these values.  Even when $n$ is small and $\rho$ is large, the approximation remains quite accurate. 

\section{Full Conditionals for Gibbs Sampler}
\setcounter{equation}{0}
\setcounter{equation}{0}

This section provides the derivations for and forms of the full conditionals for the Gibbs Sampler described in Section 3.2 of the main text. We start with the expression for the posterior distribution of $(X_1, X_2)$ given $\D = (X_1^* = x_1^*, X_2^* = x_2^*, Y_1^* = y_1^*)$. We have
\begin{align}
    \P[&X_1 = k_1, X_2 = k_2 \mid X_1^* = x_1^*, X_2^* = x_2^*, Y_1^* = y_1^*] \\
    &\propto  \P[X_1^* = x_1^*, X_2^* = x_2^*, Y_1^* = y_1^* \mid X_1 = k_1, X_2 = k_2] \nonumber \\
    &\qquad \P[X_2 = k_2 \mid X_1 = k_1] \, \P[X_1 = k_1]  \\
    &\propto \P[X_1^* = x_1^* \mid X_1 = k_1] \, \P[X_2^* = x_2^* \mid X_2 = k_2]  \nonumber \\
    &\qquad \P[Y_1^* = y_1^* \mid X_1 = k_1, X_2 = k_2] \, \P[X_2 = k_2 \mid X_1 = k_1] \, \P[X_1 = k_1] \\
    &\propto \exp\left\{-\rho_1(x_1^* - k_1)^2 \right\} \exp\left\{-\rho_2(x_2^* - k_2)^2 \right\} \nonumber \\ 
    &\qquad \cdot \exp\left\{-\frac{\rho_1}{d}(y_1^* - (k_2 - k_1))^2 \right\} \, \one[k_2 \geq k_1] \, \P[X_1 = k_1]. 
\end{align}
The full conditional for $X_1$ is then, for $k_1 \in \{x_{1,-t}, x_{1,-t}+1\}$ and $k_1 \leq k_2$,
\begin{align}
    \P[&X_1 = k_1 \mid X_2 = k_2, X_1^* = x_1^*, X_2^* = x_2^*, Y_1^* = y_1^*] \\
    &\propto \exp\left\{-\rho_1(x_1^* - k_1)^2 \right\}\exp\left\{-\frac{\rho_1}{d}(y_1^* - k_2 + k_1)^2 \right\} \, \P[X_1 = k_1] \\
    &\propto \exp\left\{-\rho_1(k_1^2 - 2k_1x_1^*) -\frac{\rho_1}{d}(k_1^2 - 2k_1(k_2 - y_1^*)) \right\} \, \P[X_1 = k_1]\\
    &\propto \exp\left\{-\frac{d+1}{d}\rho_1k_1^2 + 2\rho_1(x_1^* + \frac{1}{d}(k_2 - y_1^*))k_1 \right\} \, \P[X_1 = k_1] \\
    &\propto \exp\left\{-\frac{d+1}{d}\rho_1\left[k_1 - \frac{dx_1^* + (k_2 - y_1^*)}{d+1} \right]^2 \right\} \, \P[X_1 = k_1]. 
\end{align}
This full conditional is straightforward to sample from. 

The full conditional for $X_2$ is, for $k_2 \in \{k_1, k_1+1, \ldots\}$, 
\begin{align}
    \P[&X_2 = k_2 \mid X_1 = k_1, X_1^* = x_1^*, X_2^* = x_2^*, Y_1^* = y_1^*] \\
    &\propto \exp\left\{-\rho_2(x_2^* - k_2)^2 \right\}\exp\left\{-\frac{\rho_1}{d}(y_1^* + k_1 - k_2)^2 \right\}\\
    &\propto \exp\left\{-\rho_2(k_2^2 - 2k_2x_2^*) -\frac{\rho_1}{d}(k_2^2 - 2k_2(y_1^* + k_1)) \right\} \\
    &\propto \exp\left\{-\left(\rho_2 + \frac{\rho_1}{d}\right)k_2^2 + 2\left(\rho_2x_2^* + \frac{\rho_1}{d}(y_1^* + k_1)\right)k_2 \right\} \\
    &\propto \exp\left\{-\left(\rho_2 + \frac{\rho_1}{d}\right)\left[k_2 - \frac{\rho_2x_2^* + \frac{\rho_1}{d}(y_1^* + k_1)}{\rho_2 + \frac{\rho_1}{d}}\right]^2 \right\}.
\end{align}
This is a truncated discrete Gaussian distribution centered at $\frac{\rho_2x_2^* + \frac{\rho_1}{d}(y_1^* + k_1)}{\rho_2 + \frac{\rho_1}{d}}$.  It can be easily sampled over a grid, since the tails of the distribution decay rapidly. Using these full conditionals, the adversary can sample from the posterior distribution  and examine the marginal posterior distribution for $X_1$.

\section{Prior Sensitivity}
\setcounter{equation}{0}

\begin{figure}[p]
    \centering
    \includegraphics[width=\columnwidth]{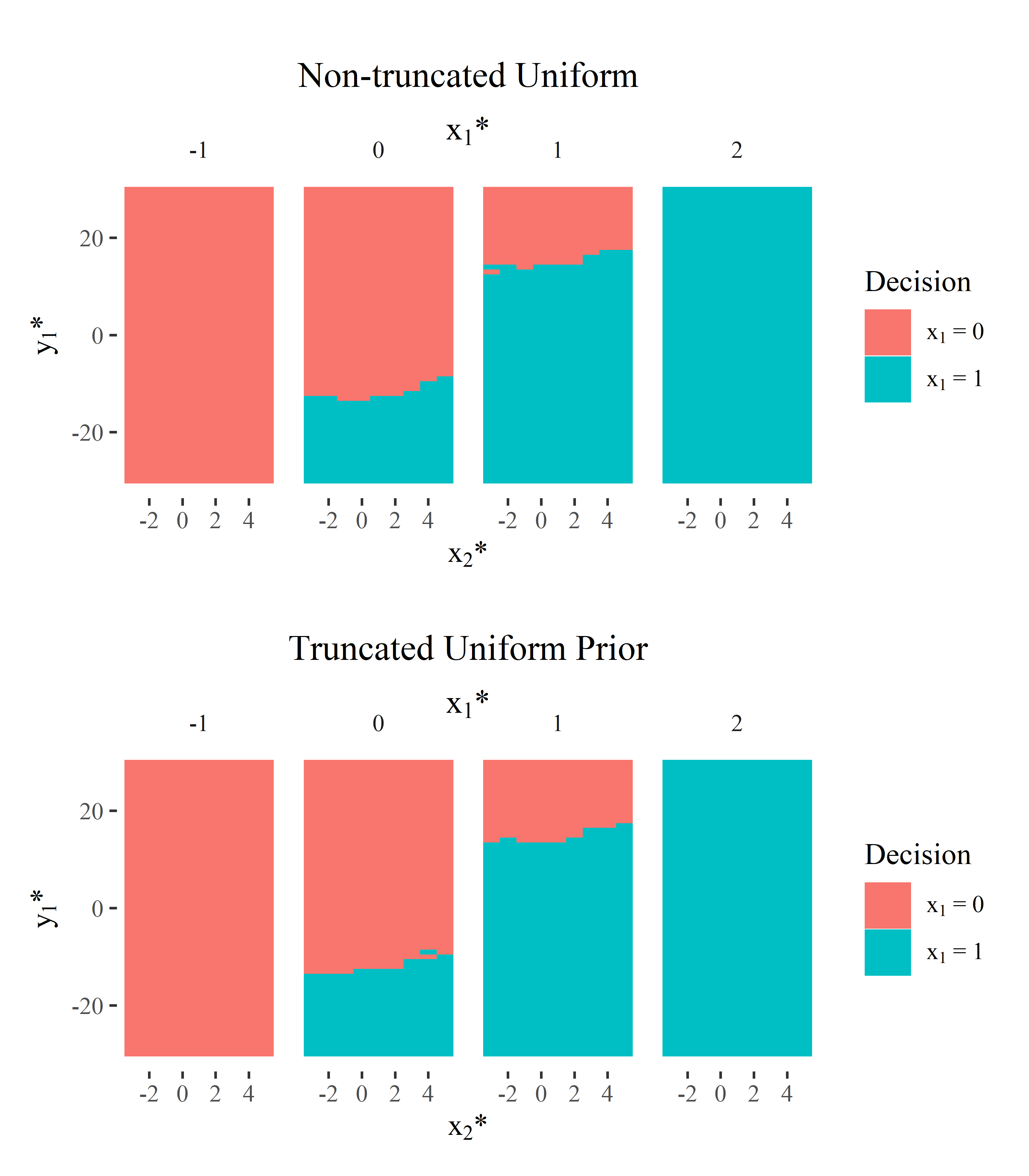}
    \caption{\linespread{1}\normalsize Adversary's decision under 0-1 loss for each combination of $x_1^*$, $x_2^*$, and $y_1^*$. The top plot reproduces Figure 4 from the main text, while the bottom plot uses the prior $X_2 \mid X_1 \sim \textsf{Unif}(\{k_1, \ldots, 10\})$. 
    Privacy parameters are set as in the census application, $p = 1/2$, and $d = 27$. $10^3$ MCMC draws are taken for each combination in most cases. When the posterior probability $X_1 = 1$ is close to 0.5, the number of MCMC draws is increased to $2.5 \times 10^5$.}
    \label{fig:plot9a}
\end{figure}

This section examines how sensitive the analysis producing Figure 4 in Section 4.3 of the main text is to the choice of the adversary's prior on $X_2 \mid X_1$. In particular, since the prior 
\begin{equation}
(X_2 \mid X_1 = k_1) \sim \textsf{Unif}(\{k_1, k_1 + 1, \ldots\}), \qquad k_1 \in \{0,1\},
\end{equation}
is an improper probability distribution with unbounded support, it may unduly favor values that are practically implausible. To determine whether this is the case, we re-do the analysis producing Figure 4 with a selection of other priors and examine how the conclusions change. We assume throughout that the prior probability for $X_1$ is $p = 1/2$, the number of other blocks is $d = 27$, and the true counts are $x_1 = x_2 = 1$. Figure 4 from the main text is reproduced as the top panel of Figure \ref{fig:plot9a}, for ease of comparison.

We begin by examining a variation on the uniform prior used in the main text.  Suppose that an adversary, utilizing information from auxiliary data sources, knows that the number of individuals in block group $g_2$ with characteristics $c$ is at most $10$. A reasonable prior might then be
\begin{equation}
    (X_2 \mid X_1 = k_1) \sim \textsf{Unif}(\{k_1, \ldots, 10\}), \qquad k_1 \in \{0,1\}.
\end{equation}
This prior has bounded support and does not place any prior probability on very extreme values for $X_2$. The results for this prior are presented on the bottom panel of Figure \ref{fig:plot9a}. We do not observe a substantial change between the truncated and non-truncated priors; for both, the adversary makes the correct decision 59\% of the time. The lack of change is likely due to the fact that, as suggested by Table 8 in the main text, the unbounded uniform prior allows the data to rule out implausible values away from $x_2$.

\begin{figure}[p]
    \centering
    \includegraphics[width=\columnwidth]{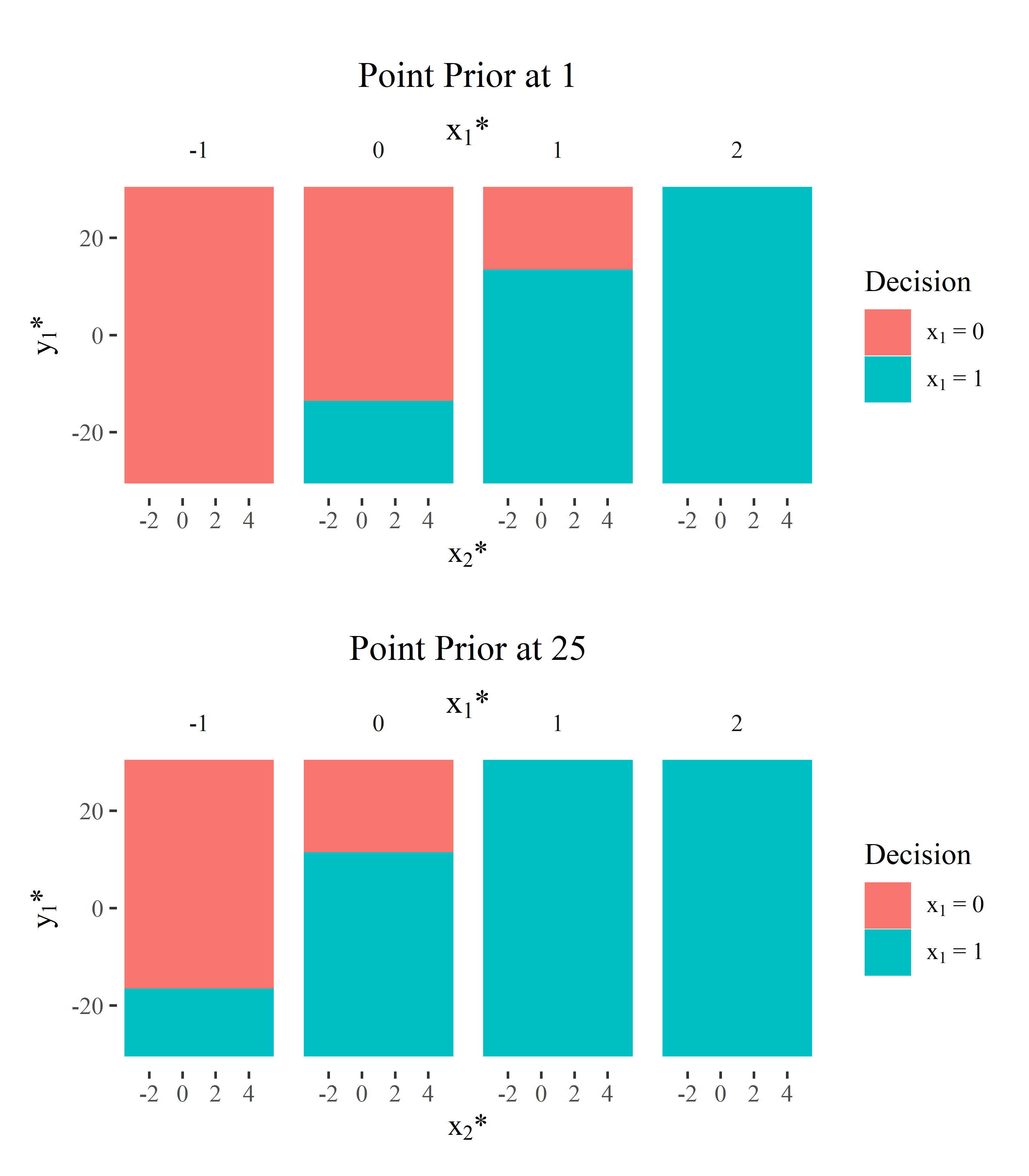}
    \caption{\linespread{1}\normalsize Adversary's decision under 0-1 loss for each combination of $x_1^*$, $x_2^*$, and $y_1^*$. Priors are of the form $\P[X_2 = k_2 \mid X_1 = k_1] = 1$ for $k_2 = 1$ (top) and $k_2 = 25$ (bottom). Privacy parameters are set as in the census application, $p = 1/2$, and $d = 27$. $10^3$ MCMC draws are taken for each combination in most cases. When the posterior probability $X_1 = 1$ is close to 0.5, the number of MCMC draws is increased to $10^6$.}
    \label{fig:plot9b}
\end{figure}

Another interesting comparison is to the case where the adversary knows a priori that $x_2 = 1$. This corresponds to a prior with all the probability mass on $X_2 = 1$. As a consequence, the adversary's  decision about $x_1$ does not depend on $x_2^*$.  The results for this prior are presented in the top panel of Figure \ref{fig:plot9b}.  We do not observe a substantial change from the previous two figures; the adversary still makes the correct decision 59\% of the time, even with perfect knowledge at the second level. The agreement between this result and the uniform priors suggests that the uniform priors are not biasing the results to any substantial degree.

One might take the above as evidence that the choice of prior for $X_2 \mid X_1$ is of little importance. We demonstrate that this is not the case by examining the results under a poorly specified prior. Suppose that the adversary incorrectly believes that $x_2 = 25$ and places a prior with all the probability mass on $X_2 = 25$. The results for this prior are presented in the bottom panel of Figure \ref{fig:plot9b}. We observe a substantial change between this plot and the previous three: the adversary now makes the correct decision 74\% of the time. But consider the counterfactual where in truth $x_1 = 0$ (and $x_2 = 0$). Now the misspecified prior leads the adversary astray, and they make the correct decision only 42\% of the time (the distribution of $X_1^*$ and $X_2^*$ change in the counterfactual, so the probability is not simply $100\% - 74\%$). Evidently, an inaccurate prior can impact the results, possibly to the detriment or benefit of the adversary depending on the value of $x_1$. Of course, in practical contexts the adversary does not know whether they benefit or suffer from an informative prior.  Given that the uniform prior (with support that includes the true count) allows the distributions of the noisy counts to fully determine the posterior probability computations, it appears to be a sensible choice when evaluating statistical disclosure risks.

\section{Composition of Risk}
\setcounter{equation}{0}

In this section, we briefly examine how the risk measures from the main text behave under composition. That is, if the Census Bureau were to perform a second data release, how would the risks from the two releases combine? Let $X_{1i}^*$ be the released noisy count from the $i^{th}$ release and $x_{1i}^*$ be the corresponding observed value. Recall that  the disclosure risk from the first release is
\begin{align}
    R'(x_{11}^*) = \frac{\P[X_1 = x_1 \mid X_{11}^* = x_{11}^*]}{\P[X_1 = x_1]}.
\end{align}
We can similarly examine the disclosure risk from the second release. Assuming the releases are sequential, the adversary will have already observed $x_{11}^*$, so their prior probability for the second release corresponds exactly to their posterior from the first release. That is,
\begin{align}
    R'(x_{12}^* \mid x_{11}^*) = \frac{\P[X_1 = x_1 \mid X_{12}^* = x_{12}^*, X_{11}^* = x_{11}^*]}{\P[X_1 = x_1 \mid X_{11}^* = x_{11}^*]}.
\end{align}
This quantity is analogous to $R'(x_{11}^*)$ and, in practice, the posterior in the numerator can be decomposed as follows via Bayes Theorem:
\begin{align}
    \P[&X_1 = x_1 \mid X_{12}^* = x_{12}^*, X_{11}^* = x_{11}^*] \nonumber \\
    &= \frac{\P[X_{12}^* = x_{12}^* \mid X_1 = x_1, X_{11}^* = x_{11}^*] \, \P[X_1 = x_1 \mid X_{11}^* = x_{11}^*]}{\P[X_{12}^* = x_{12}^* \mid X_{11}^* = x_{11}^*]} \\
    &= \frac{\P[X_{12}^* = x_{12}^* \mid X_1 = x_1] \, \P[X_1 = x_1 \mid X_{11}^* = x_{11}^*]}{\sum_{k_1 = x_{1,-t}}^{x_{1,-t}+1}\P[X_{12}^* = x_{12}^* \mid X_1 = k_1] \, \P[X_1 = k_1 \mid X_{11}^* = x_{11}^*]}. \label{eq:seq_bayes}
\end{align}
The latter equality assumes that the mechanism for releasing $X_{12}^*$ does not depend on the observed $x_{11}^*$. (\ref{eq:seq_bayes}) has a form identical to the form of $\P[X_1 = x_1 \mid X_{1}^* = x_1^*]$ in the main text, except that the prior is conditional on the observed $x_{11}^*$ from the first release. This means that the analysis of the second release can proceed exactly as the first with the only difference being an ``updated" prior.

The total risk from the two releases is then
\begin{align}
    R'(x_{11}^*, x_{12}^*) &= \frac{\P[X_1 = x_1 \mid X_{11}^* = x_{11}^*, X_{12}^* = x_{12}^*]}{\P[X_1 = x_1]} \\
    &= \frac{\P[X_1 = x_1 \mid X_{12}^* = x_{12}^*, X_{11}^* = x_{11}^*]}{\P[X_1 = x_1 \mid X_{11}^* = x_{11}^*]} \cdot \frac{\P[X_1 = x_1 \mid X_{11}^* = x_{11}^*]}{\P[X_1 = x_1]} \\
    &= R'(x_{12}^* \mid x_{11}^*) R'(x_{11}^*).
\end{align}
This argument generalizes to an arbitrary number of releases. Letting $m$ be the total number of releases, the total risk composes as
\begin{align}
    R'(x_{11}^*, \ldots, x_{1m}^*) = R'(x_{1m}^* \mid x_{11}^*, \ldots, x_{1,m-1}^*) \cdots R'(x_{11}^*).
\end{align}
Thus, the cumulative risk is simply the product of the risk from each release. The generalized marginal risk and generalized probability the adversary makes the correct decision are straightforward to compute from the generalized $R'$.

\section{The Effect of Post-Processing}
\setcounter{equation}{0}

In this section, we illustrate how a post-processing step could affect the risk analysis in this article. Our intent is not to give a complete treatment of this but rather to provide a rough intuition. Thus, in this analysis, we make a substantial number of simplifying assumptions about the adversary and the way the post-processing is performed compared to the TopDown algorithm used for the 2020 decennial census data.

To begin, we outline our illustrative post-processing algorithm. Let $\tX_1,\tX_2$ be the post-processed counts corresponding to $X_1,X_2$ and $\tx_1,\tx_2$ be their observed values. Similarly, let $\tY_1^{(1)},\ldots,\tY_1^{(d)}$ be the post-processed counts corresponding to $Y_1^{(1)},\ldots, Y_1^{(d)}$ and $\ty_1^{(1)},\ldots,\ty_1^{(d)}$ be their observed values. We define the post-processing algorithm at the block level as follows. Taking $\tx_2$ as fixed, we enforce the aggregation constraint $\tx_2 = \tx_1 + \sum_{i=1}^d \ty_1^{(i)}$, while minimizing the sum of squared deviations from the noisy counts:
\begin{equation}
    \argmin_{\tx_1, \ty_1^{(1)}, \ldots, \ty_1^{(d)}} \bigg\{(\tx_1 - x_1^*)^2 + \sum_{i=1}^d(\ty_1^{(i)} - y_1^{(i)*})^2 \bigg\}.
\end{equation}
The post-processing algorithm used in the TopDown algorithm enforces several aggregation constraints and minimizes a weighted sum of squared deviations involving more quantities, so this is a substantial simplification, but one that we expect to roughly approximate the effects of the true algorithm. Letting $x_1^*$ and $y_1^*$ be the observed noisy counts corresponding to $X_1$ and $Y_1$, this simplified problem has a closed form solution, which we denote $\bar{x}_1$:
\begin{equation}
    \bar{x}_1 = \frac{dx_1^* + (\tx_2 - y_1^*)}{d + 1}.
\end{equation}
It is possible for $\bar{x}_1$ to be outside the range $[0, \tx_2]$ or to be non-integer valued. To correct for this, we truncate the solution to be in the correct range and round to the nearest integer. The complete post-processing algorithm includes a non-negativity constraint in the optimization and performs a second controlled rounding step, although we expect this change to have a limited effect for our illustration. We denote the final solution to the optimization as $f(x_1^*, y_1^*, \tx_2)$, which is given by
\begin{equation}
    f(x_1^*, y_1^*, \tx_2) = 
    \begin{cases}
        0, & \textrm{if } \bar{x}_1 < 0; \\
        \tx_2, & \textrm{if } \bar{x}_1 > \tx_2; \\
        [\bar{x}_1], & \textrm{otherwise.}
    \end{cases}
\end{equation}

We now return to the perspective of the adversary. From the above, the likelihood from the post-processing step is simply an indicator variable
\begin{align}
    \P[\tX_1 = \tx_1 \mid X_1^* = x_1^*, Y_1^* = y_1^*, \tX_2 = \tx_2] = \one[\tx_1 = f(x_1^*, y_1^*, \tx_2)].
\end{align}
The full likelihood is then, assuming that the adversary considers $\tx_1,\tx_2$ and not $\ty_1^{(1)}, \ldots, \ty_1^{(d)}$,
\begin{align}
    \P&[\tX_1 = \tx_1 \mid X_1 = k_1, \tX_2 = \tx_2] \nonumber \\
    &= \sum_{x_1^* = -\infty}^\infty\sum_{y_1^* = -\infty}^\infty \P[\tX_1 = \tx_1 \mid X_1^* = x_1^*, Y_1^* = y_1^*, \tX_2 = \tx_2] \nonumber \\
    & \hspace{82pt} \P[X_1^* = x_1^* \mid X_1 = k_1] \, \P[Y_1^* = y_1^* \mid X_1 = k_1] \\
    &= \sum_{x_1^* = -\infty}^\infty\sum_{y_1^* = -\infty}^\infty \one[\tx_1 = f(x_1^*, y_1^*, \tx_2)] \, \P[X_1^* = x_1^* \mid X_1 = k_1] \nonumber \\
    & \hspace{82pt} \P[Y_1^* = y_1^* \mid X_1 = k_1].
\end{align}
To simplify $\P[Y_1^* = y_1^* \mid X_1 = k_1]$, we assume the adversary knows $x_2$ exactly a priori, in addition to making the approximation from Section \ref{app:DG_approx}. Assuming as in the main text that the true $x_1 = x_2 = 1$, the known count is $x_{1,-t} = 0$, and the adversary's prior on $X_1$ is Bernoulli with parameter $p$, the posterior probability the adversary makes the correct decision is
\begin{align}
    &\P[X_1 = 1 \mid \tX_1 = \tx_1, \tX_2 = \tX_2] \nonumber \\
    &= \frac{\P[\tX_1 = \tx_1 \mid X_1 = 1, \tX_2 = \tx_2] \, p}{\P[\tX_1 = \tx_1 \mid X_1 = 1, \tX_2 = \tx_2] \, p + \P[\tX_1 = \tx_1 \mid X_1 = 0, \tX_2 = \tx_2] (1-p)}.
\end{align}
Finally, for comparison to the results from the main text, we can marginalize out $\tX_1$ from the posterior:
\begin{align} \label{eq:marg_risk}
    \P[X_1 = 1 \mid x_1 = 1, \tX_2 = \tx_2] = \sum_{\tx_1 = 0}^{\tx_2} &\P[X_1 = 1 \mid \tX_1 = \tx_1, \tX_2 = \tx_2] \nonumber \\
     &\P[\tX_1 = \tx_1 \mid x_1 = 1, \tX_2 = \tx_2].
\end{align}
Note that the result will vary with $\tx_2$.

\begin{figure}[t]
    \centering
    \includegraphics{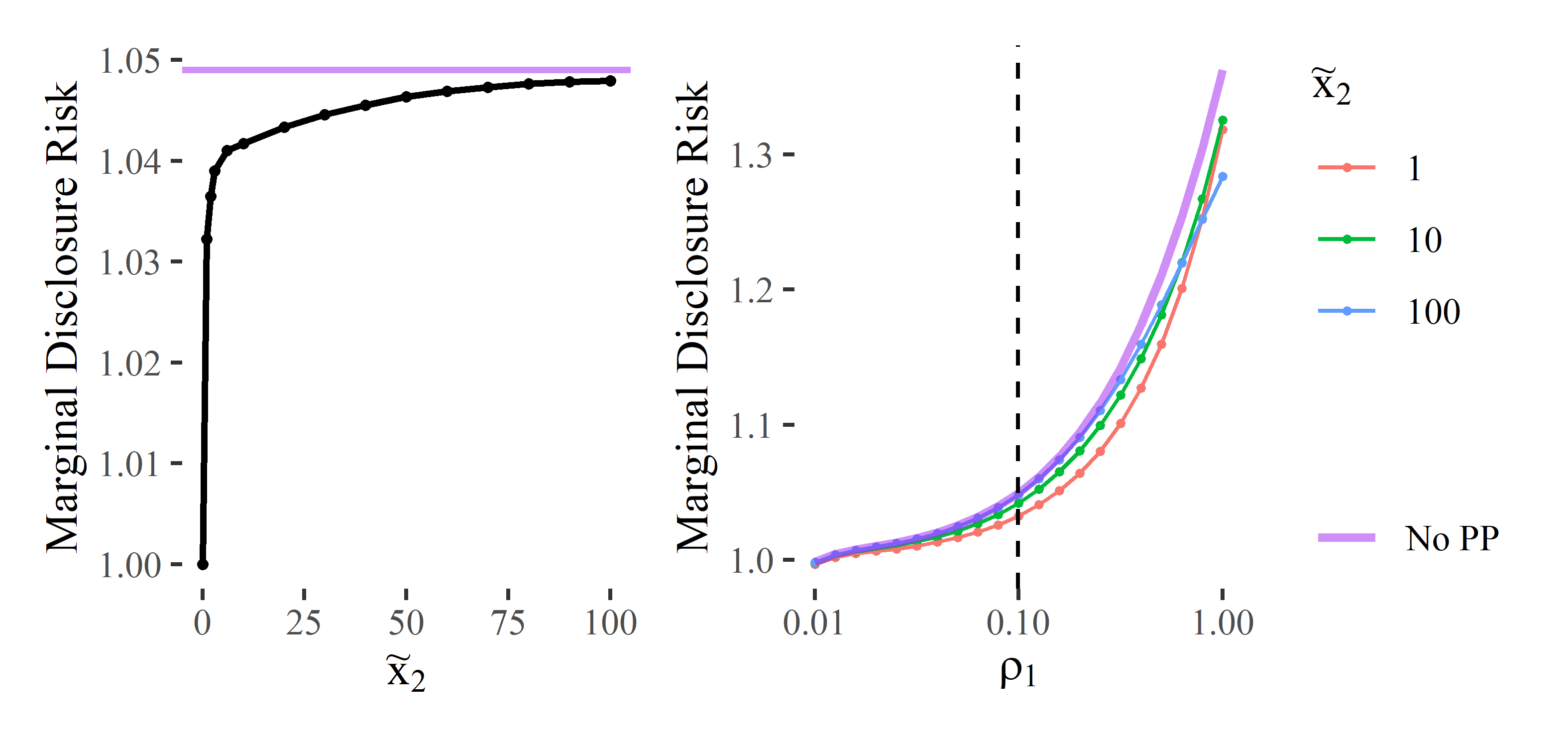}
    \caption{The left panel plots the marginal disclosure risk from (\ref{eq:marg_risk}) as a function of $\tx_2$ when $\rho_1 \approx 0.099$. The right panel plots the marginal disclosure risk from (\ref{eq:marg_risk}) as a function of $\rho_1$ colored by $\tx_2 \in \{1, 10, 100\}$. The dashed line represents $\rho_1 \approx 0.099$. In both panels, the purple line represents the marginal disclosure risk without post-processing. Both set $p = 1/2$ and assume the adversary knows that $x_2 = 1$.}
    \label{fig:plot10}
\end{figure}

We  now examine whether, on average, releasing the counts with post-processing will have lower disclosure risk than releasing the counts without post-processing. The first panel of Figure \ref{fig:plot10} compares the marginal disclosure risks in the case where $\rho_1 \approx 0.099$ for various values of $\tx_2$. We find that the marginal risk with post-processing is bounded above by the marginal risk without post-processing, as expected. Larger values of $\tx_2$ give risks closer to the bound, which makes sense intuitively; larger values of $\tx_2$ allow for a larger range of possible observed $\tx_1$, which will make it easier for the adversary to ``work backward" to $x_1^*$. The second panel of Figure \ref{fig:plot10} compares the marginal disclosure risks as a function of $\rho_1$ for a selection of $\tx_2$. We observe a similar effect, with the marginal risk without post-processing providing an upper bound on the marginal risk with post-processing. In general, we find that the bound is fairly tight; the reduction in disclosure risk due to the post-processing is minor (given the simplifications and assumptions we make).

\end{document}